\documentclass[a4paper,11pt]{article}
\usepackage{jheppub} 
\usepackage{lineno}
\usepackage{amssymb, amsmath}
\usepackage{listings}
\usepackage{multirow}
\usepackage{makecell}
\usepackage{array}
\usepackage{subfigure}
\usepackage{siunitx} 
\usepackage{booktabs}

\usepackage{slashed}
\usepackage{comment}
\usepackage{enumitem}
\newlist{baseline}{enumerate}{1}
\setlist[baseline]{label=\textbf{(\arabic*)},ref=\arabic*,
  leftmargin=1.8em,labelsep=0.5em,itemsep=2pt,topsep=4pt}

\arxivnumber{2512.21558} 

\title{Deep-learning jet flavor tagging for precision hadronic Higgs measurements at future $e^+e^-$ Higgs factories}

\author[a,c,1]{Xinzhu Wang,}

\author[b,c,1]{Yifan Zhu,}

\author[a,c,1]{Chunxiang Zhu,}

\author[d,e]{Jianfeng Jiang,}

\author[d,e]{Manqi Ruan,}

\author[f,*]{Kun Wang,}

\author[a,b,c,*]{Haijun Yang,}

\author[d,e]{Yongfeng Zhu}

\note[*]{Corresponding author.}

\note[1]{These authors contributed equally to the work.}

\affiliation[a]{Institute of Nuclear and Particle Physics, School of Physics and Astronomy, Shanghai Jiao Tong University, Shanghai 200240, China}

\affiliation[b]{Tsung-Dao Lee Institute, Shanghai Jiao Tong University, Shanghai 201210, China}

\affiliation[c]{State Key Laboratory of Dark Matter Physics, Shanghai Key Laboratory for Particle Physics and Cosmology, Key Laboratory for Particle Astrophysics and Cosmology (MoE), Shanghai 200240, China}

\affiliation[d]{Institute of High Energy Physics, Chinese Academy of Sciences, Beijing 100049, China}

\affiliation[e]{University of Chinese Academy of Sciences, 19A Yuquan Road, Beijing 100049, China}

\affiliation[f]{School of Physics, Faculty of Basic Sciences, University of Shanghai for Science and Technology, Shanghai 200093, China}

\emailAdd{kwang@usst.edu.cn}

\emailAdd{haijun.yang@sjtu.edu.cn}

\abstract{Precise measurements of Higgs decays into quarks and gluons are essential for probing the Yukawa couplings of the Higgs boson and testing the flavor structure of the Standard Model. We investigate the process $e^+e^- \to ZH$ at $\sqrt{s}=240~\mathrm{GeV}$ at a future $e^+e^-$ Higgs factory, taking the CEPC design as a benchmark. The analysis focuses on events with $Z\to\nu\bar\nu$ and hadronic Higgs decays $H\to b\bar b$, $c\bar c$, $s\bar s$ and $gg$. Jet flavor is identified using state-of-the-art particle-level deep neural network taggers (ParticleNet, Particle Transformer and More-Interaction Particle Transformer), whose per-jet outputs are combined with global event observables in a two-stage analysis employing XGBoost classifiers to separate the four Higgs decay modes from the dominant two- and four-fermion Standard Model backgrounds. Assuming an integrated luminosity of $20\,\mathrm{ab}^{-1}$, we obtain projected relative precision on $\sigma(ZH)\times\mathrm{Br}(H\to X)$ of 0.17\% for $X=b\bar b$, 1.06\% for $c\bar c$, 0.50\% for $gg$ and 68\% for $s\bar s$.  Compared with the CEPC published results, the precisions for $H\to c\bar c$ and $H\to gg$ are improved by about $43\%$ and $29\%$, respectively.  For $H\to s\bar s$ we present a quantitative sensitivity estimation corresponding to a statistical significance of about $1.5\sigma$. These results highlight the potential of deep-learning-based jet flavor tagging for precision studies of Higgs decays at future $e^+e^-$ Higgs factories.
}

\begin{document}
\maketitle
\flushbottom

\section{\label{sec:Introduction}Introduction}

The discovery of the Higgs boson at the Large Hadron Collider (LHC) by the ATLAS and CMS Collaborations~\cite{ATLAS:Higgs_comb,CMS:Higgs_comb} completed the particle content predicted by the Standard Model (SM) and confirmed the mechanism of electroweak symmetry breaking~\cite{novaes2000standardmodelintroduction, Higgs:1964ia1, Englert:1964et, Guralnik:1964eu, Kibble:1967sv}.  
Over the past decade, a broad program of precision measurements has examined the Higgs couplings to weak gauge bosons and third-generation fermions through various decay channels, including 
$H\to WW/ZZ$~\cite{CMS:Higgs_comb,ATLAS:HyyHZZ,ATLAS:HWW},
$H\to \gamma\gamma$~\cite{CMS:Hyy,ATLAS:HyyHZZ}, 
$H\to b\bar{b}$~\cite{CMS:Hbb,ATLAS:VHbb}, and 
$H\to \tau^+\tau^-$~\cite{CMS:Htautau,ATLAS:Htautau}. 
These results are in excellent agreement with SM predictions within the current experimental uncertainties.  
However, some crucial aspects of the Higgs sector remain unexplored, particularly its interactions with second-generation fermions such as $H\to s\bar{s}$.

In the SM, the Yukawa couplings of the Higgs to second-generation fermions are suppressed relative to the third generation, leading to rare decay modes such as 
$H\to c\bar{c}$, $H\to s\bar{s}$, and $H\to \mu^+\mu^-$.  
While emerging evidence for $H\to c\bar{c}$~\cite{ATLAS:VHbbcc} and $H\to \mu^+\mu^-$~\cite{ATLAS:2025coj, CMS:2020xwi} has been reported at the LHC, direct measurements of the light quark channels remain challenging.  
This is primarily due to the overwhelming quantum chromodynamics (QCD) background in hadronic environments~\cite{Skands:2012ts}, where jets originating from $c$ and $s$ quarks must be distinguished from a large number of jets initiated by $u/d$ quarks and gluons, which exhibit similar substructure.  
Nevertheless, precision measurements of these rare decays are essential not only to examine the Higgs mechanism but also to test the SM to its limits and probe possible extensions, such as models featuring modified Yukawa hierarchies or flavor-violating Higgs interactions~\cite{DAmbrosio:2002vsn, Branco:2011iw, Redi:2011zi, Harnik:2012pb}.

Future electron-positron ($e^+e^-$) colliders, such as the Circular Electron-Positron Collider (CEPC)~\cite{CEPCStudyGroup:2018rmc,CEPCStudyGroup:2018ghi,CEPCStudyGroup:2023quu,CEPCStudyGroup:2025kmw}, the Future Circular Collider-ee (FCC-ee)~\cite{agapov2022} and the International Linear Collider (ILC)~\cite{BARISH_2013}, are designed to operate as high-luminosity `Higgs factories'.  
These colliders offer a clean experimental environment with precise initial-state control, low pileup and minimal QCD backgrounds, making them ideal for precision Higgs measurements that are extremely difficult at hadron colliders.  
Specifically, the $e^+e^- \to ZH$ process with $Z$ decaying into $\nu\bar{\nu}$ provides a clean final-state topology, where the Higgs boson recoils against the missing momentum, allowing easier isolation of hadronic Higgs decays with minimal background contamination.
While this work uses the CEPC detector and luminosity assumptions as a benchmark, the developed analysis methodology is broadly applicable to other future $e^+e^-$ Higgs factories.

A key challenge in studying hadronic Higgs decays is jet flavor tagging. In the $ZH$ events considered in this paper, hadronic Higgs decays are typically reconstructed as a two-jet final state, with the jet angular separation depending on the Higgs kinematics.
The flavors, or the origins of these jets, are essential to distinguishing between different Higgs decay modes, such as $H\to b\bar{b}$, $c\bar{c}$, $s\bar{s}$ and $gg$, and to separating them from background processes.
At the LHC, modern machine learning algorithms like GN2~\cite{ATLAS:2025dkv} and DeepJet~\cite{Bols:2020bkb} have achieved excellent performance in identifying $b$- and $c$-jets, but they struggle to separate light-flavor jets ($s$-, $u$-, $d$-jets and gluon jets), which typically lack distinctive displaced vertices.
This limitation presents a significant obstacle to accurately measuring $H\to s\bar{s}$ and to improving the precision for $H\to c\bar{c}$ and $H\to gg$.
While complementary approaches, such as dihadron fragmentation, have been proposed to probe the $u$- and $d$-quark Yukawa sectors~\cite{Cao:2025wfg}, dedicated studies of strange-jet tagging at $e^+e^-$ colliders already exist~\cite{Duarte-Campderros:2018ouv,Nakai:2020kuu}. Here we focus instead on a full hadronic Higgs analysis in a two-stage framework.

To address these challenges, current state-of-the-art deep learning models, such as ParticleNet (PN)~\cite{Qu_2020}, Particle Transformer (ParT)~\cite{qu2024particletransformerjettagging}, and More-Interaction Particle Transformer (MIParT)~\cite{JOI:MIParT_2025}, operate directly on the particle-level information of each jet. By considering the relationships between particles within each jet, these newer models are able to capture more complex patterns of interaction, which is crucial for distinguishing jets originating from light-flavor quarks even in the presence of significant QCD backgrounds. 
To further improve signal-background separation, we combine the outputs of these three deep learning models with an event-level classifier, XGBoost~\cite{Roe:2004, Yang:2005, XGBoost}.  
This two-stage classification approach allows us to effectively separate Higgs decays from backgrounds by exploiting both jet-level information and global event kinematics.  
The purpose of this design is not only to improve discrimination, but also to preserve explicit jet-flavor information at the event level, which is important for interpretability and future precision studies.

In this work, we develop and validate a measurement-oriented two-stage analysis strategy for hadronic Higgs decays at a future lepton collider. In addition to providing a systematic sensitivity estimate for $H\to s\bar{s}$ and updated precision projections for other hadronic Higgs decays, we also examine the complementarity of different jet taggers and the physical interpretation of the event-level classifier.
We show that, with the proposed tagging methods, the projected precision for $\sigma(ZH) \times \text{Br}(H\to c\bar{c})$ and $\sigma(ZH) \times \text{Br}(H\to gg)$ improves by about 43\% and 29\%, respectively, compared with the CEPC published results~\cite{An_2019}.  
For $H\to s\bar{s}$, we
provide a benchmark sensitivity of $1.5\sigma$ for single $ZH\to \nu\nu s\bar{s}$ channel, marking a significant step towards probing the strange Yukawa coupling.

This paper is organized as follows.
In Sec.~\ref{sec:method} we describe the collider and detector setup, event generation, jet reconstruction and flavor tagging, as well as the event selection and multivariate analysis strategy.
Section~\ref{sec:results} presents the jet- and event-level performance and the projected sensitivities to 
$\sigma(ZH)\times\mathrm{Br}(H\to b\bar b,\, c\bar c,\, gg,\, s\bar s)$, 
including a comparison with the CEPC Higgs benchmark~\cite{An_2019} and FCC-ee projections~\cite{DelVecchio:2025gzw}.
We conclude and discuss future prospects in Sec.~\ref{sec:conclusion}.

\section{Simulation and analysis strategy}
\label{sec:method}

In this section, we describe the collider and detector assumptions, the Monte~Carlo (MC) simulation, jet reconstruction and flavor tagging, and the event-level classification strategy used to obtain the projections presented in the following.  We take the CEPC as a concrete benchmark of a high-luminosity Higgs factory, but the analysis chain is otherwise generic and can be straightforwardly reinterpreted for other future $e^+e^-$ colliders.

\subsection{Simulation setup and event samples}
\label{sec:simsetup}

The CEPC is designed to operate at $\sqrt{s}=240~\text{GeV}$ and to collect an integrated luminosity of $20\,\text{ab}^{-1}$ in Higgs factory mode, corresponding to four million Higgs bosons produced in association with a $Z$ boson~\cite{CEPCStudyGroup:2025kmw, Yang:2025wew,Ai__2025}.  The reference detector features a low-mass silicon vertex detector and tracker~\cite{li2024beamtestbaselinevertex}, surrounded by highly granular electromagnetic and hadronic calorimeters optimized for particle-flow reconstruction~\cite{song2024studyresidualartificialneural}. This configuration provides excellent impact-parameter resolution, jet-energy and angular resolution, and particle identification capabilities, which are all critical for heavy- and light-flavor jet tagging. 

Hard-scattering events are generated with \textsc{MadGraph}~3.6.3~\cite{Alwall:2014hca} and interfaced with \textsc{Pythia}~8.313~\cite{Bierlich:2022pfr} for parton showering and hadronization. Initial-state radiation is included in the event generation, and QCD shower radiation is enabled in \textsc{Pythia}~8.313. Detector effects are simulated with \textsc{Delphes3}~\cite{deFavereau:2013fsa} using the official CEPC detector card, which parametrises the tracking efficiency, momentum and impact-parameter resolutions, calorimeter responses and the performance of particle-flow reconstruction.  Unless stated otherwise, all projections are normalized to an integrated luminosity of $20~\text{ab}^{-1}$ at $\sqrt{s}=240~\text{GeV}$.

\begin{figure*}[htbp]
    \centering
    \begin{tabular}{cc}
        \includegraphics[width=0.35\textwidth]{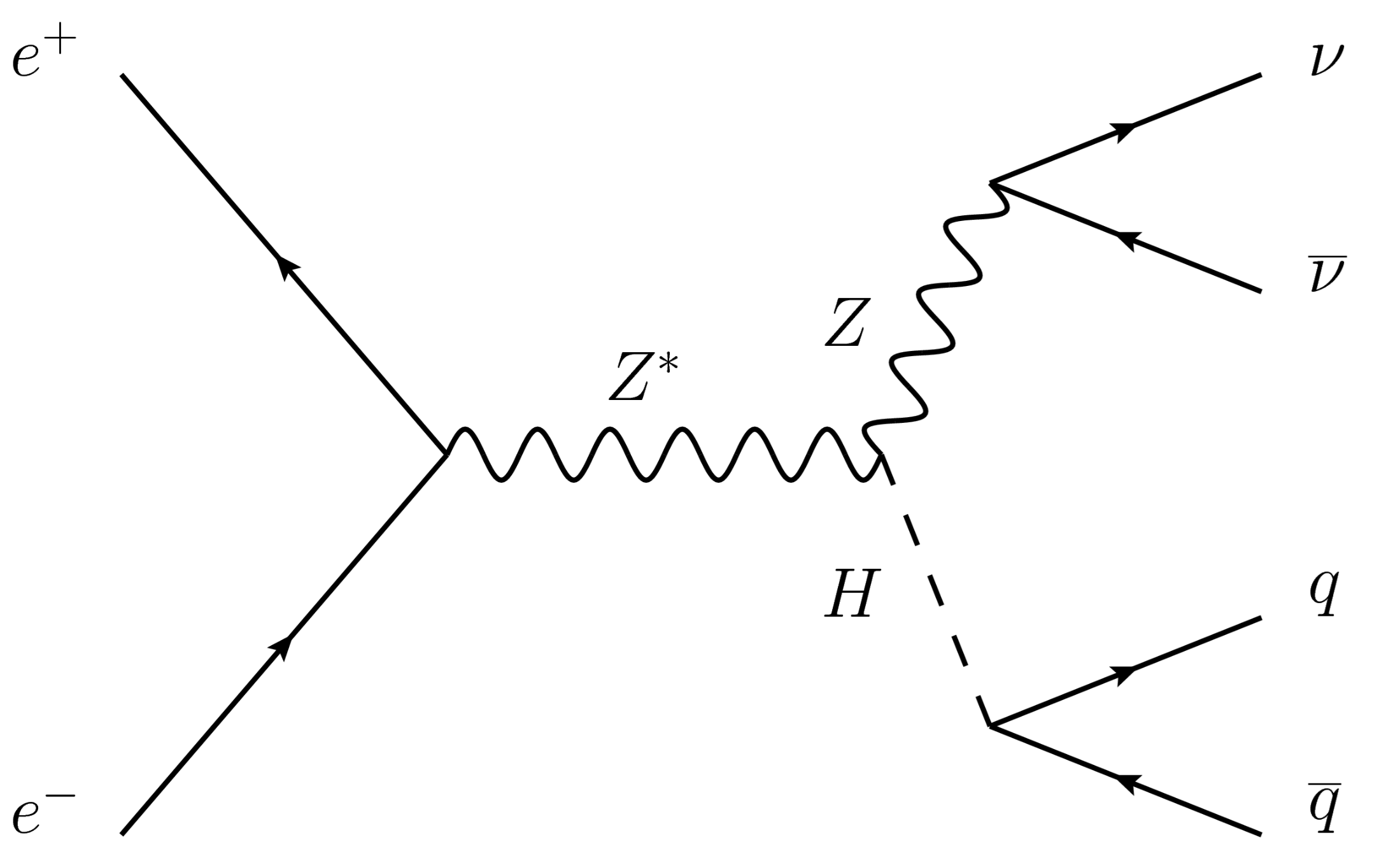} &
        \includegraphics[width=0.35\textwidth]{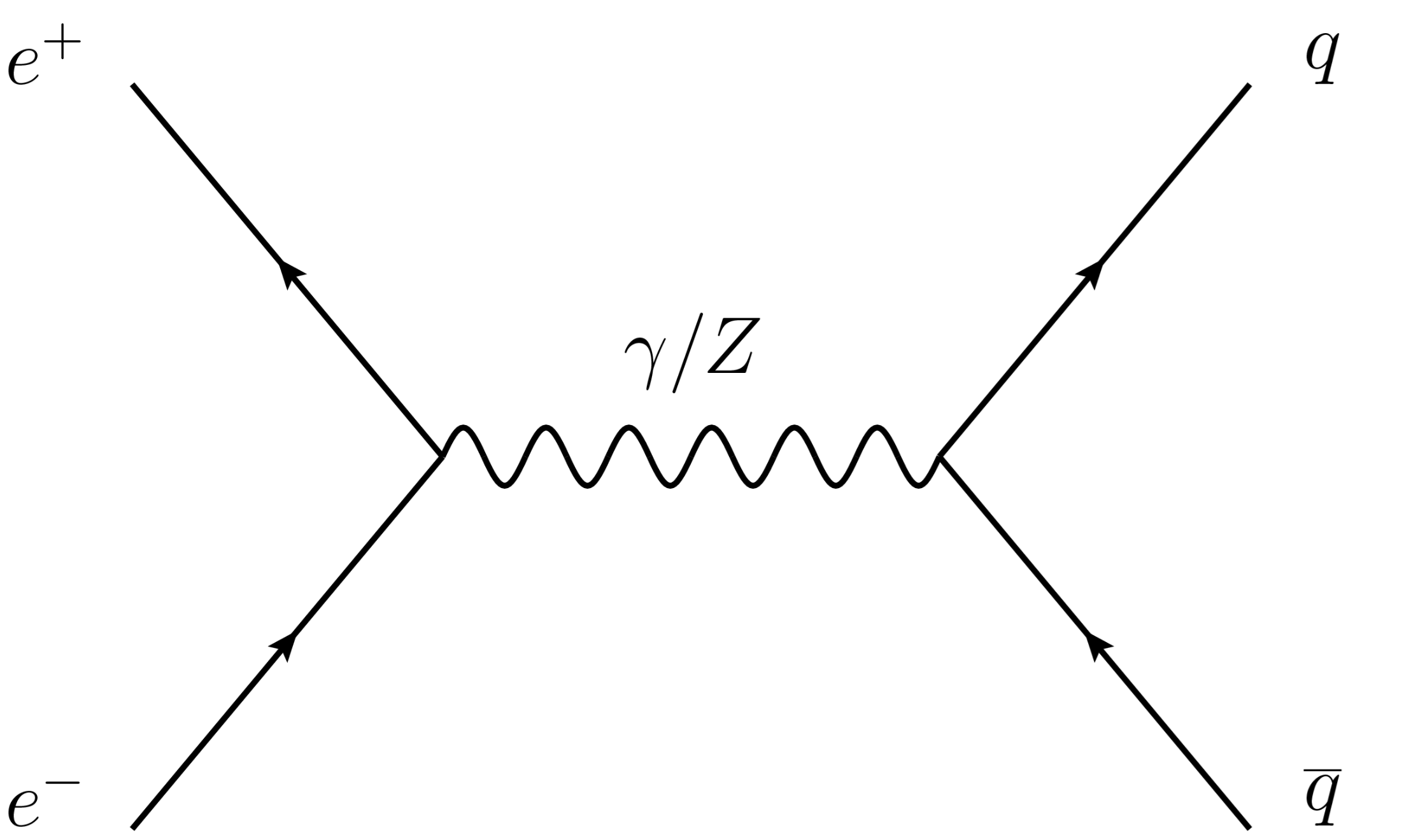} \\
        \textbf{(a)} &
        \textbf{(b)} \\[1em]
        \multicolumn{2}{c}{
            \includegraphics[width=0.70\textwidth]{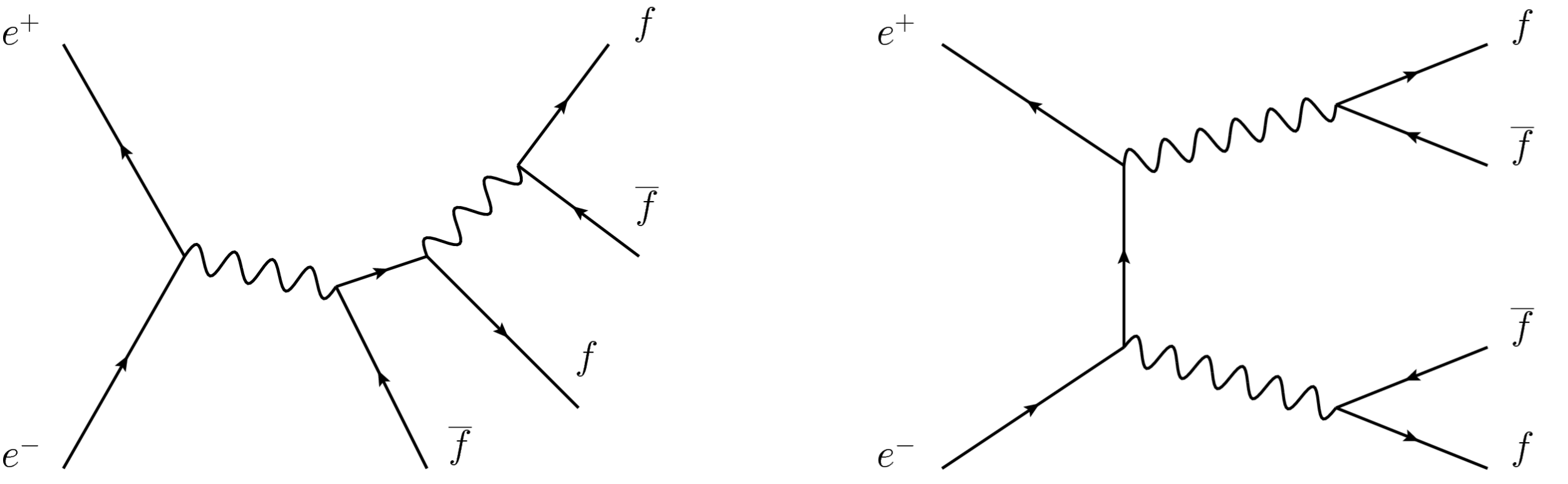}
        } \\
        \multicolumn{2}{c}{\textbf{(c)}} \\
    \end{tabular}

    \caption{
        Representative Feynman diagrams for the signal and background processes:
        (a) $e^+e^- \to ZH$ with $Z \to \nu\bar{\nu}$ and $H \to q\bar{q}$;  
        (b) the two-fermion background $e^+e^- \to q\bar{q}$;  
        (c) the representative four-fermion background $e^+e^- \to \nu\bar{\nu}q\bar{q}$.
    }
    \label{fig:feynman_overview}
\end{figure*}


The signal process considered in this work is the associated Higgs production
\begin{equation}
  e^+e^- \to ZH \,, \quad Z\to\nu\bar{\nu}\,, \quad
  H \to q\bar q \text{ or } H\to gg\,,
\end{equation}
where the Higgs boson decays hadronically and the $Z$ boson decays invisibly.  
We explicitly simulate the Higgs decay modes $H\to b\bar b$, $c\bar c$, $s\bar s$, $gg$, $u\bar u$, and $d\bar d$, with the production cross sections and decay branching ratios taken from Refs.~\cite{Zhu_2022,bagnaschi2025higgsbosondecaysupdates}.  
The dominant non-Higgs backgrounds are two-fermion ($2f$) and four-fermion ($4f$) processes, in particular $e^+e^-\to q\bar q$ and $e^+e^-\to\nu\bar{\nu}q\bar q$. 
We also include the Higgs-associated background where the Higgs boson decays to a pair of Z bosons, which then decay invisibly and hadronically, respectively. 
Representative Feynman diagrams for the signal and the main background topologies are shown in Fig.~\ref{fig:feynman_overview}.

\begin{table*}[htbp]
  \centering
    \caption{
Cross sections and simulated events for signal and background processes. Here $\sigma$ denotes the cross section, $N_{\mathrm{exp}}$ is the expected number of events at an integrated luminosity of $20~\mathrm{ab}^{-1}$, and $N_{\mathrm{MC}}$ is the number of generated Monte Carlo events.
For all signal samples the production mode is $e^+e^- \to ZH$ with $Z\to\nu\bar\nu$; 
major backgrounds include four-fermion and two-fermion processes. 
For the two-fermion samples, the quoted cross sections are evaluated after the generator-level event filter with an efficiency of about 2\%.
    }

  \label{tab:sim_samples}
  \vspace{0.2cm}
  \begin{tabular}{c c c c c}
    \toprule
  \textbf{Category} & \textbf{Channel} &
  $\boldsymbol{\sigma}$ [fb] &
  $\boldsymbol{N_\text{exp}}$ [k] &
  $\boldsymbol{N_\text{MC}}$ [k] \\
    \midrule
    \multirow{6}{*}{Signal}
      & $e^+e^- \to ZH, Z \to\nu\bar\nu, H\to b\bar b$ & 26.71              & 534.2   & 10000 \\
      & $e^+e^- \to ZH, Z \to\nu\bar\nu, H\to c\bar c$ & 1.35               & 27.0    & 10000 \\
      & $e^+e^- \to ZH, Z \to\nu\bar\nu, H\to s\bar s$ & 0.01               & 0.2     & 10000 \\
      & $e^+e^- \to ZH, Z \to\nu\bar\nu, H\to gg$      & 3.97               & 79.4    & 10000 \\
      & $e^+e^- \to ZH, Z \to\nu\bar\nu, H\to u\bar u$ & $2.77\times10^{-5}$& 0.00056 & 10000 \\
      & $e^+e^- \to ZH, Z \to\nu\bar\nu, H\to d\bar d$ & $8.00\times10^{-3}$& 0.16    & 10000 \\
    \midrule
    \multirow{3}{*}{Background}
      & $e^+e^- \to \nu\bar{\nu}H,\,\, H\to ZZ^* \to\nu\bar{\nu}q\bar{q}$ & 0.34   & 6.8     & 10000 \\
      & $e^+e^- \to 4f$                              & 369.71 & 7394.2  &  40000 \\
      & $e^+e^- \to 2f$                              &  1320 & 26400.0 &  25000  \\
    \bottomrule
  \end{tabular}
\end{table*}

The MC samples used in this analysis are summarized in Table~\ref{tab:sim_samples}.
For each Higgs decay mode we generate $10^7$ events, corresponding to negligible MC statistical uncertainty in the subsequent training of the jet taggers.
For the $2f$ processes, generator-level phase-space cuts ($110 < m_{q\bar{q}} < 140$~GeV and $15 < p_\textup{T}(q) < 100$~GeV) are applied to enhance the acceptance in the kinematic region relevant for the $ZH(\nu\bar{\nu})$ topology, resulting in a selection efficiency of about 2\% applied to the cross section quoted in the Table~\ref{tab:sim_samples}.
The channels $H\to u\bar u$ and $H\to d\bar d$ have cross sections several orders of magnitude smaller than $H\to s\bar s$ and are therefore not included in the event-level signal extraction; however, they are kept in the jet-tagging training in order to stabilize the light-flavor classification.  Contributions from $H\to WW^*$ are effectively suppressed by the lepton veto used in the event selection and are neglected in this analysis, consistent with Ref.~\cite{CPC_2019_Higgs}. Contributions from $H\to\tau^+\tau^-$ are also neglected. Although $\mathrm{Br}(H\to\tau^+\tau^-)\approx 3\%$, requiring both $\tau$ leptons to be misidentified as jets (at the $\sim1\%$ level per $\tau$) leads to an overall suppression of $\mathcal{O}(10^{-4})$, making this contribution negligible compared to $H\to s\bar{s}$~\cite{Giagu:2022tauID}.

\subsection{Two-stage jet and event-level analysis}
\label{sec:twostage}

Detector effects are simulated with \textsc{Delphes3}~\cite{deFavereau:2013fsa}, in which jet clustering is performed with \textsc{FastJet}~\cite{Cacciari:2011ma}. Jets are reconstructed from all stable visible particles, using the $e^{+}e^{-}$-$k_t$ algorithm~\cite{CATANI1991432} in the exclusive two-jet mode, following the CEPC baseline strategy.
Each event contains exactly two reconstructed jets, which are taken as Higgs decay products and are used as inputs to the jet flavor taggers.

A central ingredient of the analysis is the identification of the jet flavor originating from the Higgs decay.  Instead of relying on hand-crafted high-level observables, we employ three state-of-the-art particle-level deep neural networks: PN, ParT and MIParT. Each jet is represented as an unordered set of reconstructed particles, characterized by per-particle kinematics, particle-identification (PID) flags and track-trajectory information as listed in Table~\ref{tab:tag_input_features}.  PN is based on dynamic graph convolutions and is particularly effective at learning local substructure features; ParT employs transformer-style attention to capture long-range correlations between particles within jets; MIParT simplifies the attention mechanism to increase efficiency while retaining strong representational power for particle interactions.
For training purposes, each jet is assigned one of 11 Monte Carlo truth labels: $b,\bar b,c,\bar c,s,\bar s,u,\bar u,d,\bar d$, or $g$. The quark--antiquark separation has little impact on the final projected sensitivities of this work. For clarity, quark and antiquark categories are combined in some figures and discussions. The performance of the taggers is quantified using multi-class confusion matrices, which will be discussed in Sec.~\ref{sec:results}.

\begin{table*}[htbp]
    \centering
    \caption{Input feature categories used for training the jet-tagging networks. Kinematic features ($\Delta\eta$, $\Delta\phi$, $\log p_{\mathrm{T}}$, etc.) are defined relative to the jet axis; particle identification features are binary flags indicating the reconstructed particle type; trajectory features ($\tanh d_{0}$, $\tanh d_{z}$, etc.) encode impact parameter information for displaced vertex sensitivity.}
    \label{tab:tag_input_features}
    \setlength{\tabcolsep}{20.0pt}
    \vspace{0.2cm}
    \begin{tabular}{@{\hspace{50pt}} lll @{\hspace{50pt}}}
        \toprule
        \textbf{Kinematics} & \textbf{Particle identification} & \textbf{Trajectory} \\
        \midrule
        $\Delta \eta$ & charged kaon & $\tanh d_{0}$ \\
        $\Delta \phi$ & pion & $\tanh d_{z}$ \\
        $\log p_{\mathrm{T}}$ & charged hadron & $\tanh \sigma_{d_{0}}$ \\
        $\log E$ & neutral hadron & $\tanh \sigma_{d_{z}}$ \\
        $\log (p_{\mathrm{T}} / p_{\mathrm{T}}^{\text{(jet)}})$ & electron &  \\
        $\log (E / E^{\text{(jet)}})$ & muon &  \\
        $\Delta R$ & photon &  \\
        \bottomrule
    \end{tabular}
\end{table*}

The input features used for jet tagging are summarized in Table~\ref{tab:tag_input_features}.  They consist of: (i)~particle-level kinematic quantities, defined relative to the jet axis and normalized by the jet transverse momentum or energy; (ii)~binary PID indicators provided by \texttt{Delphes3}; and (iii)~impact-parameter related observables for charged tracks, which encode information on displaced vertices. 
Before being fed into the networks, all features are standardized using the mean and standard deviation computed from the training sample.
For each jet tagger, we train three independent models with different random initializations, in order to assess the stability of the results.  
Approximately $25\%$ of the signal events are reserved for the development and evaluation of the jet tagging networks, 
while the remaining $75\%$ are kept for the event-level analysis described below. 
Background samples are excluded from the jet tagging procedure.

\begin{figure*}[htbp]
    \centering
    \includegraphics[width=0.95\linewidth]{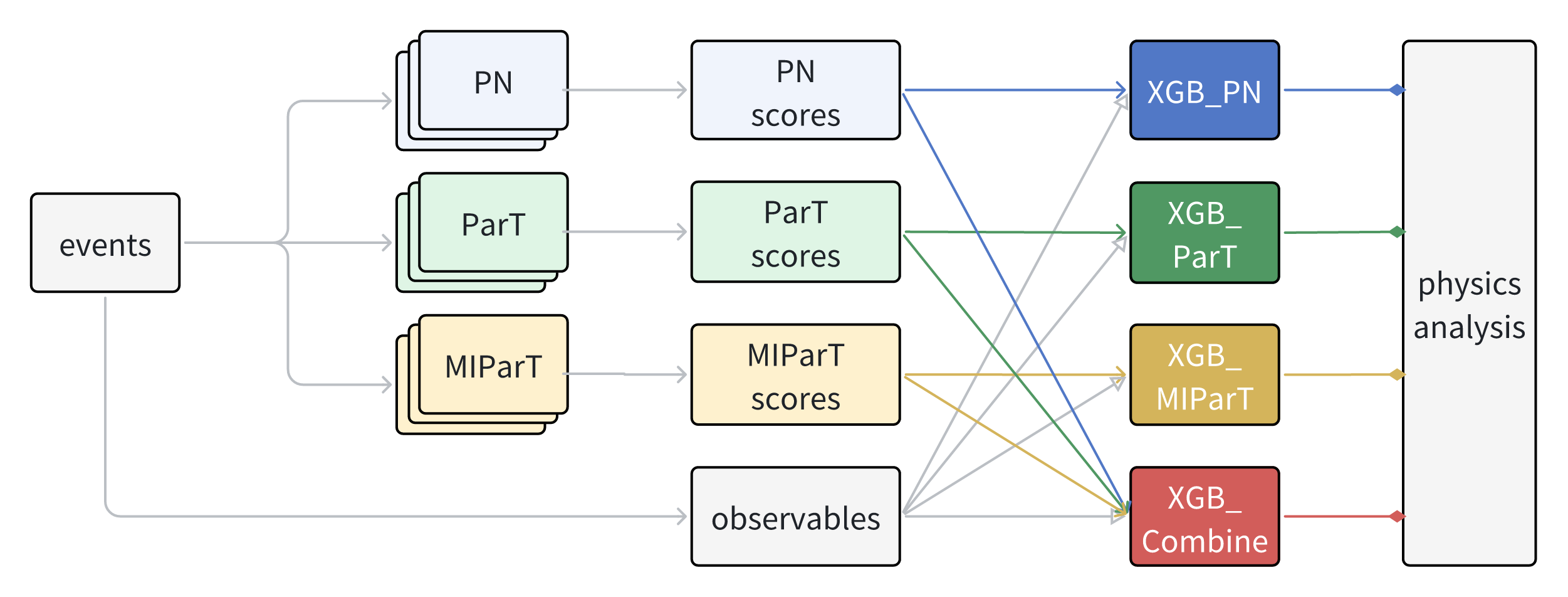}
    \caption{
        Schematic overview of the analysis strategy.
        Events are first processed by the three jet-level taggers (PN, ParT,
        MIParT).  Their per-jet flavor scores, together with global event
        observables, are then used as inputs to a set of XGBoost classifiers:
        three using a single tagger each (\texttt{XGB\_PN},
        \texttt{XGB\_ParT}, \texttt{XGB\_MIParT}) and one combined classifier
        (\texttt{XGB\_Combined}) that combines all three taggers simultaneously.
            }
    \label{fig:_analysis_strategy}
\end{figure*}

The overall analysis follows a hierarchical two-stage design, illustrated schematically in Fig.~\ref{fig:_analysis_strategy}. 
At the jet level, three deep-learning jet taggers, PN, ParT and MIParT, are trained to predict per-jet flavor labels, each repeated three times using different training initializations to ensure statistical robustness. 
These jet-level outputs, together with global event observables, serve as inputs to subsequent event-level classifiers. This modular structure decouples microscopic jet-substructure learning from event-level correlations, enhancing both sensitivity and interpretability.

On top of the jet-level flavor information, we exploit global event kinematics to enhance the separation between different Higgs decay modes and backgrounds.  After reconstructing the jets and evaluating the PN, ParT and MIParT outputs for the two Higgs-candidate jets, we build a set of event-level observables including:
\begin{itemize}
  \item Single-jet variables: Transverse momentum $p_{\mathrm{T}}$, longitudinal momentum $p_z$, pseudorapidity $\eta$, polar angle $\theta$, azimuthal angle $\phi$, and energy $E$ of the leading and subleading jets.
  \item Dijet system variables: Invariant mass and angular separations ($\Delta\eta_{jj}$, $\Delta\phi_{jj}$, $\Delta\theta_{jj}$) of the jet pair;
  \item Missing energy (ME) related variables: The transverse and longitudinal components of the missing energy (MET and MEZ) and its polar and azimuthal angles ($\eta_\mathrm{ME}$, $\theta_\mathrm{ME}$ and $\phi_\mathrm{ME}$) and the ratio between MET and the scalar sum of $p_\textup{T}$ of the jets ($\text{MET}/H_\textup{T}$).
  \item Correlations between each jet and MET, including $\Delta\eta(j,\ \text{MET})$, $\Delta\phi(j,\ \text{MET})$ and $\Delta\theta(j,\ \text{MET})$;
  \item The jet flavor tagging outputs of PN, ParT and MIParT for each of the jets, which summarize the microscopic substructure information learned at the jet level.
\end{itemize}

A representative list of these observables is given in Table~\ref{tab:BDT_inputs}.  
A quantitative study of the relative importance of these input variables in this analysis, based on feature importance and SHAP\cite{lundberg2017unifiedapproachinterpretingmodel} values, is presented in Appendix~\ref{app:XGB:importance_shap}.
Before multivariate training, a loose preselection is applied to enhance the $ZH(\nu\bar\nu)$ topology while keeping high signal efficiency.  In particular, we require the leading jet to satisfy $p_\textup{T} \in (15,100)~\text{GeV}$ and $p_{z} \in (-95,95)~\text{GeV}$, the missing longitudinal energy to be within $E_{z,\text{miss}} \in (-55,55)~\text{GeV}$, and the dijet invariant mass to lie in the window $m_{jj} \in (110,140)~\text{GeV}$. 
After this combined preselection, the low-$E_{\text{miss}}$ region is already strongly suppressed, so the subsequent requirements on $p_z$ and $E_{z,\text{miss}}$ are applied only as symmetric ranges, without introducing additional lower thresholds.
These requirements retain about $80\%$ of the Higgs signal while rejecting more than $97\%$ of the inclusive $2f$ and $4f$ backgrounds.
The corresponding kinematic distributions for signal and background processes, and their physical interpretation, are discussed in Appendix~\ref{app:preselection}.

\begin{table*}[htbp]
    \centering
    \caption{Inputs for XGBoost classification.\label{tab:BDT_inputs}}
    \setlength{\tabcolsep}{15.0pt}
    \vspace{0.2cm}
    \begin{tabular}{@{\hspace{10pt}} l l @{\hspace{10pt}}}
        \toprule
        \textbf{Category} & \textbf{Features} \\
        \midrule
        \text{Single-jet kinematics} & $p_\textup{T}$, $p_z$, $\eta$, $\theta$, $\phi$, energy \\
        \text{Jet shape and composition} & $n_\mathrm{particles}$, $\Delta R$, $\Delta p_\textup{T}$ \\
        \text{Dijet kinematics} & $m_{jj}$, $\Delta\eta_{jj}$, $\Delta\theta_{jj}$, $\Delta\phi_{jj}$ \\
        \text{Missing energy} & MET, MEZ, $\eta_\mathrm{ME}$, $\theta_\mathrm{ME}$, $\phi_\mathrm{ME}$, $\mathrm{MET}/H_\textup{T}$ \\
        \text{Jet–MET correlations} & $\Delta\eta$, $\Delta\phi$, $\Delta\theta$ between each jet and MET \\
        \text{Jet flavor tagging scores} & outputs from ParticleNet, ParT, and MIParT \\
        \bottomrule
    \end{tabular}
\end{table*}

Event-level classification is performed using the XGBoost algorithm~\cite{XGBoost}, a gradient-boosted decision-tree method~\cite{Roe:2004,Yang:2005} well suited for heterogeneous inputs and for combining deep-learning–based jet information with classical event-level kinematic observables. We train six-class classifiers to distinguish $H\to b\bar b$, $H\to c\bar c$, $H\to s\bar s$, $H\to gg$, and the $2f$ and $4f$ backgrounds. All classifiers make use of the same set of event-level kinematic observables. The three baseline models differ only in the source of jet-level information, each using the per-jet scores from a single tagger (\texttt{XGB\_PN}, \texttt{XGB\_ParT}, and \texttt{XGB\_MIParT}). A fourth classifier, \texttt{XGB\_Combined}, is trained using the full collection of jet scores from all three taggers together, in addition to the same kinematic observables, allowing XGBoost to exploit the complementary information among them. For each classifier, the event-level samples are randomly split into statistically independent training, validation and test sets in a 6:2:2 ratio. Hyperparameters such as the maximum tree depth, learning rate and number of boosting rounds are tuned on the validation sample to optimize the multi-class log-loss and to avoid overfitting.  The resulting classifier outputs are used in Sec.~\ref{sec:results} to define signal-enriched regions for each Higgs decay channel and to extract the projected measurement precisions.

For the purpose of signal extraction, the output score of a given XGBoost classifier is treated as a one-dimensional discriminant for the corresponding Higgs decay hypothesis.  For each channel, all other Higgs decays as well as the $2f$ and $4f$ processes are considered as background.  After normalizing the samples to $20~\text{ab}^{-1}$, we scan over possible thresholds on the classifier score and, for each threshold, compute the expected numbers of signal ($S$) and background ($B$) events.  The statistical significance is evaluated using the profile-likelihood approximation
\begin{equation}
  Z = \sqrt{2\left[(S+B)\ln\left(1+\frac{S}{B}\right)-S\right]}\,,
  \label{eq:significance}
\end{equation}
and the optimal working point is chosen as the threshold that maximizes $Z$.  The corresponding efficiencies and event yields serve as the basis for the precision and sensitivity estimates presented in Sec.~\ref{sec:results}.

\section{Results and discussion}
\label{sec:results}

In this section we present the performance of the jet-level flavor taggers and of the event-level classifiers, and derive the expected sensitivities to hadronic Higgs decays at the CEPC Higgs factory.  Unless stated otherwise, all results are normalized to an integrated luminosity of $20~\text{ab}^{-1}$ at $\sqrt{s}=240~\text{GeV}$.

\subsection{Jet-level performance}
\label{sec:results_jet}

\begin{figure*}[htbp]
    \centering
        {
        \centering
        \begin{tabular}{ccc} 
            \includegraphics[width=0.31\textwidth]{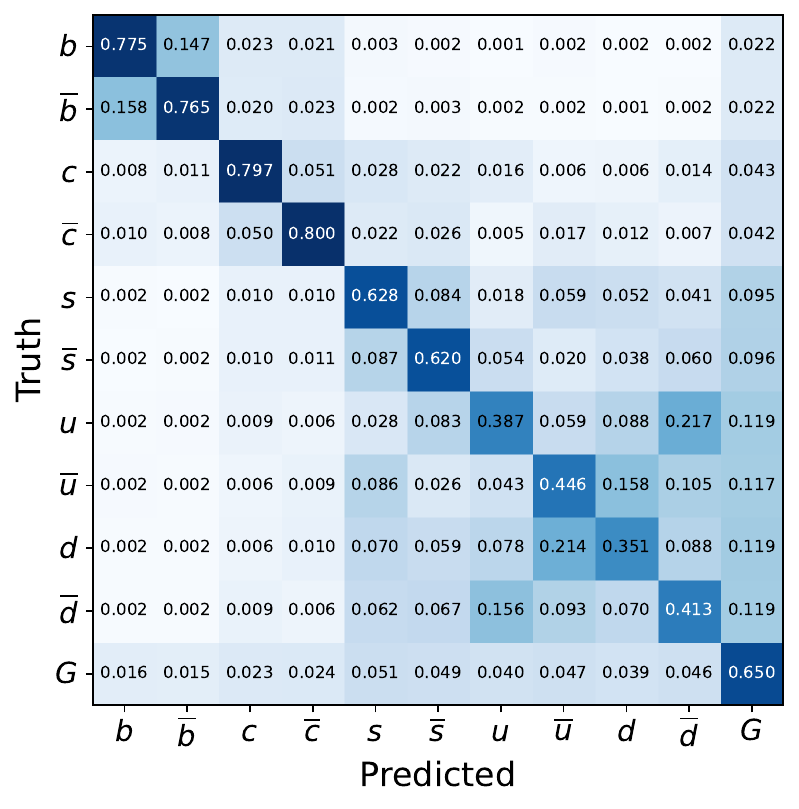} &
            \includegraphics[width=0.31\textwidth]{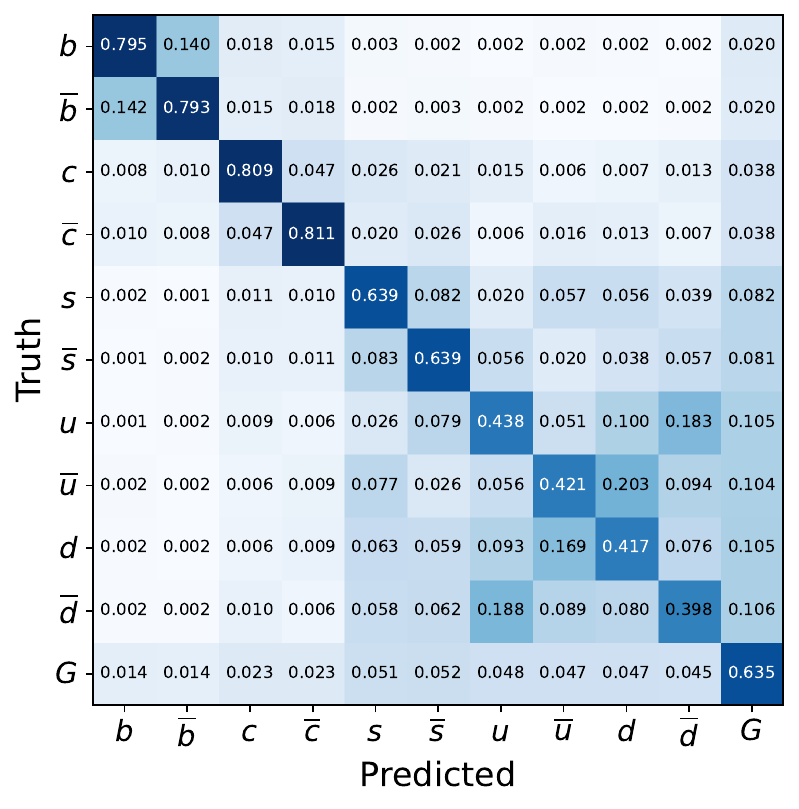} &
            \includegraphics[width=0.31\textwidth]{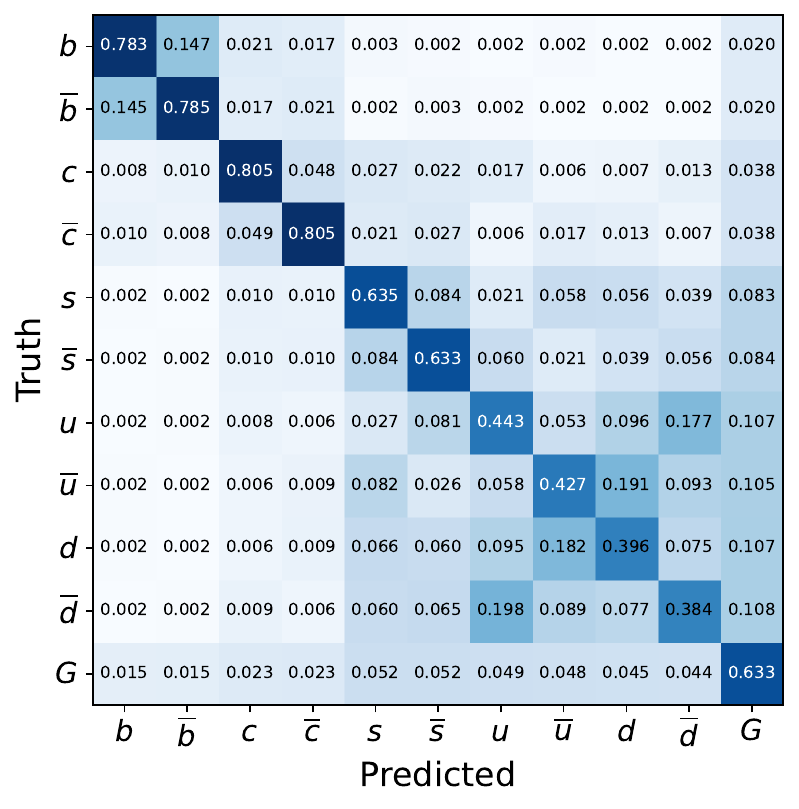} \\
            \textbf{(a)} ParticleNet &
            \textbf{(b)} ParT &
            \textbf{(c)} MIParT \\
        \end{tabular}
    } 

    \caption{
        Jet identification confusion matrices for $\nu\bar{\nu}H$,
        $H\rightarrow q\bar{q}$ events from ParticleNet (a), ParT (b) and
        MIParT (c).  Each entry gives the fraction of jets of true flavor
        $i$ classified as flavor $j$.
    }
    \label{fig:JOI_CM}
\end{figure*}

The jet flavor taggers described in Sec.~\ref{sec:twostage} are evaluated on an independent sample of $\nu\bar\nu H$ events, with $H\to q\bar q$ or $H\to gg$.
Fig.~\ref{fig:JOI_CM} shows the resulting $11\times 11$ confusion matrices for ParticleNet, ParT and MIParT, respectively.  Each matrix element $M_{ij}$ gives the fraction of jets with true flavor $i$ that are assigned to flavor $j$ by the classifier.

All three taggers achieve a good separation between $b$- and $c$-jets and the other flavors.  When particle and antiparticle jets are not distinguished, the effective $b$-tagging and $c$-tagging efficiencies exceed $90\%$, while misidentification of light-flavor jets as heavy flavors remains at the percent level.  This confirms that the CEPC detector, together with the adopted vertexing and tracking performance, provides excellent heavy-flavor tagging capabilities.

For light-flavor jets, the task is intrinsically more challenging.  When particle and antiparticle jets are not distinguished, the self-identification probability for strange jets is around $70\%$, and gluon jets are correctly classified in about $64\%$ of the cases, with substantial migration among $s$, $u/d$ and $g$ categories. These patterns are expected, as strange and gluon jets share similar particle multiplicities and radial energy profiles, and differ only through relatively subtle fragmentation and PID information.  The three taggers yield broadly comparable performance, with ParT and MIParT providing a modest improvement in heavy- and strange-flavor tagging efficiencies relative to ParticleNet, while ParticleNet achieves slightly better performance on gluon jets.  Overall, the differences are at the level of a few percentage points and will later be exploited through an ensemble combination at the event level.

\begin{figure*}[htbp]
    \centering
    \includegraphics[width=0.55\textwidth]{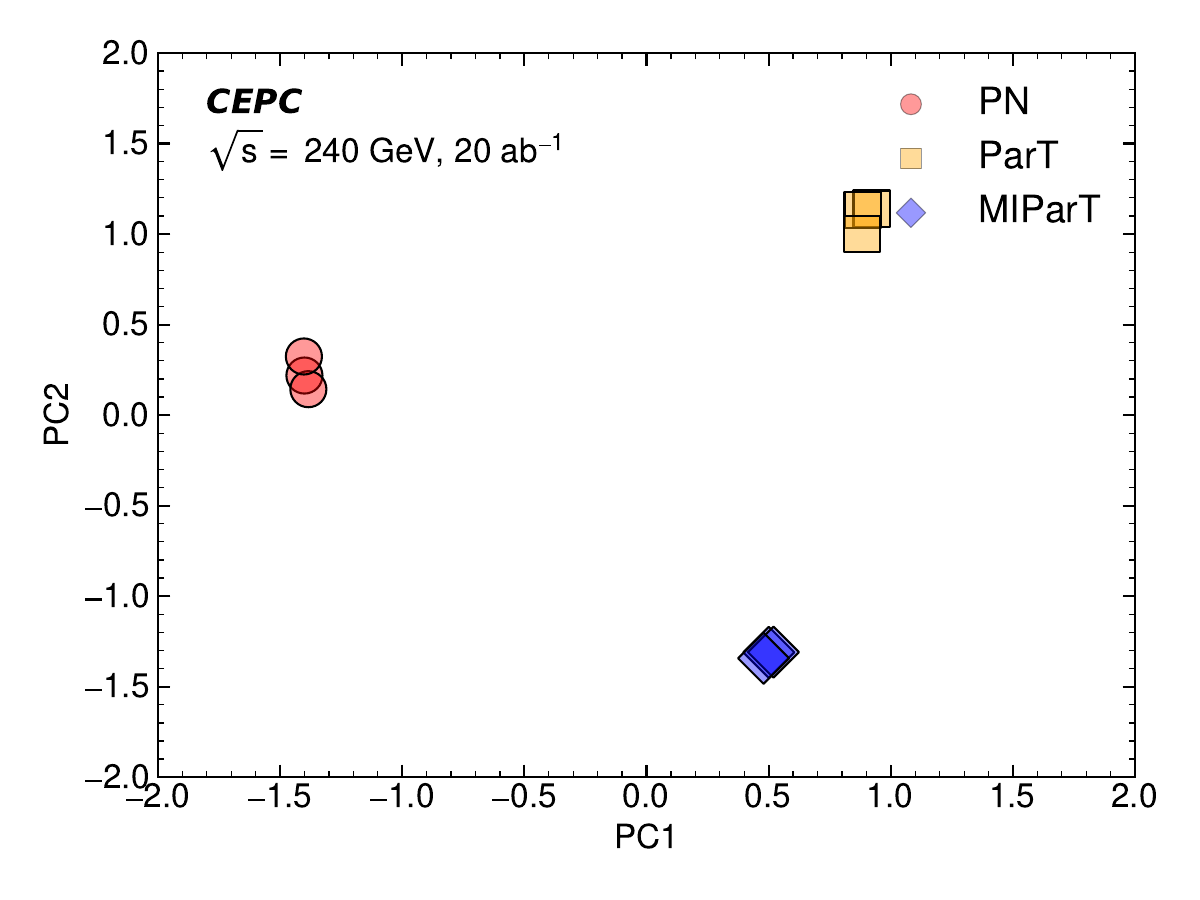}
    \caption{
    PCA projection of model-predicted $s$-jet logits for the
    $H \to s\bar{s}$ channel. Each point represents one independently
    trained model of ParticleNet (circles), ParT (squares) or MIParT
    (diamonds). Points of the same color correspond to the same model
    architecture, while the spread within a cluster reflects different
    training initializations. Models of the same architecture form compact
    clusters, indicating good training stability, while the separation
    between clusters reflects inter-architecture diversity that can be
    exploited in an ensemble.
    }

    \label{fig:pca_comparison}
\end{figure*}

To quantify the complementarity among the different taggers, we perform a principal component analysis (PCA)~\cite{PCA:2014} of their outputs at the model level. For each independently trained model, the predicted output logits for jets in the $H\to s\bar{s}$ sample are collected into a vector, and PCA is applied to these standardized model-output vectors. The first two principal components, PC1 and PC2, therefore represent the directions of largest and second-largest variation among the model outputs, rather than simple combinations of the original physics observables. The result is shown in Fig.~\ref{fig:pca_comparison}, where each point corresponds to one trained model.

Models of the same architecture cluster closely together in the PCA plane, indicating that the training is stable against random initialization. By contrast, the PN, ParT and MIParT clusters are clearly separated, showing that the three taggers retain non-identical output patterns. This inter-model diversity motivates the construction of the combined event-level classifier (\texttt{XGB\_Combined}) discussed in the next subsection.

\subsection{Event-level classification performance}
\label{sec:results_event}

\begin{figure*}[htbp]
    \centering
    \includegraphics[width=0.48\linewidth]{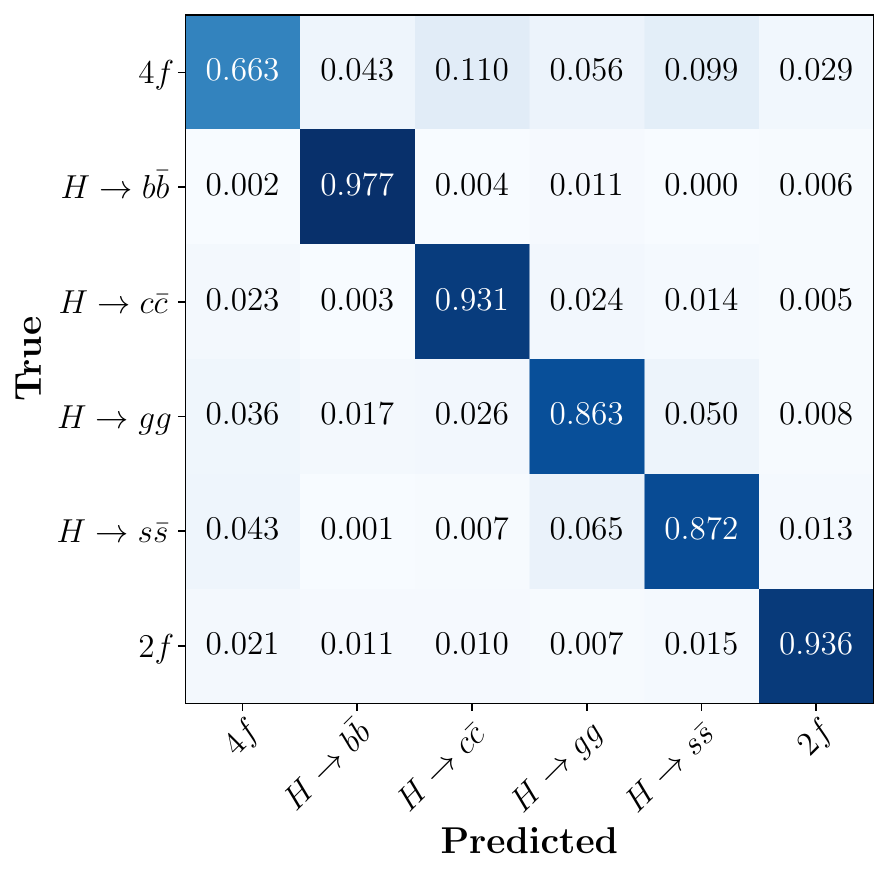}
    \hfill
    \includegraphics[width=0.49\linewidth]{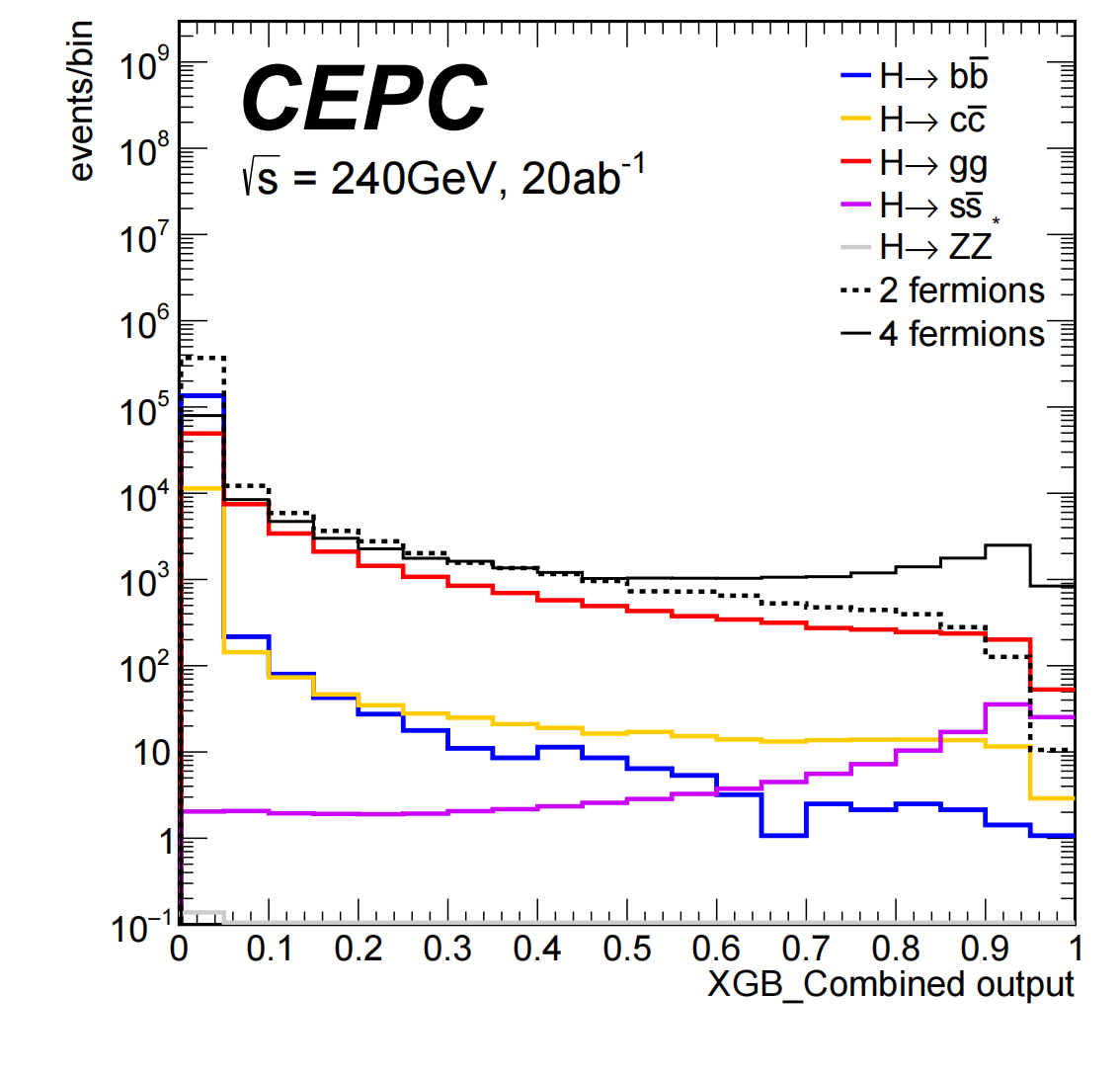}
    \caption{
        Performance of the XGBoost event classifier in the strange-enriched
        region.  Left: confusion matrix for the \texttt{XGB\_Combined}
        model, demonstrating clear separation for $H\to b\bar{b}$ and
        $H\to c\bar{c}$, while some confusion remains between $H\to s\bar{s}$
        and $H\to gg$ due to similar jet substructure.  Right: distribution
        of the \texttt{XGB\_Combined} output score, taking $H\to s\bar{s}$
        as signal, with all samples normalized to
        $20~\text{ab}^{-1}$.  Higher scores correspond to higher signal
        purity, and $4f$ processes constitute the dominant background.
    }
    \label{fig:xgb_results}
\end{figure*}

The event-level XGBoost classifiers, introduced in Sec.~\ref{sec:twostage}, are trained to separate four Higgs decay modes ($H\to b\bar b$, $c\bar c$, $s\bar s$, $gg$) from the dominant $2f$ and $4f$ backgrounds.  Fig.~\ref{fig:xgb_results} summarizes the performance of the combined classifier, \texttt{XGB\_Combined}, in the strange-enriched signal region.

The confusion matrix in the left panel shows that $H\to b\bar b$ and $H\to c\bar c$ are efficiently identified: the diagonal elements exceed $97\%$ and $93\%$ for $b\bar b$ and $c\bar c$ respectively, and the cross-contamination between these two modes is at the per-mille level.  This reflects the strong discriminating power of the heavy-flavor taggers combined with kinematic differences among the channels.

The $H\to gg$ and $H\to s\bar s$ entries have diagonal elements of about $86\%$ and $87\%$, respectively, with the largest off-diagonal migration occurring between these two channels.  This less pronounced separation is driven by the similarity of the $s$- and $g$-jet topologies.  The $2f$ and $4f$ backgrounds are efficiently controlled: the $2f$ class is identified with an accuracy of about $94\%$, and the $4f$ class with about $66\%$, the latter reflecting the broader kinematic variety of four-fermion final states.

The right panel of Fig.~\ref{fig:xgb_results} displays the distribution of the \texttt{XGB\_Combined} score when $H\to s\bar s$ is taken as the signal hypothesis.  The signal events populate the high-score region, while the $4f$ background dominates at intermediate scores and the inclusive $2f$ background is largely confined to low scores.  This behaviour confirms that the classifier successfully leverages both jet-level flavor information and global event kinematics to construct a discriminant that is monotonic in signal purity.

\subsection{Physics interpretation of the event-level classifier}
\label{sec:results_physics_interp}

For precision measurements, it is important to verify that the multivariate classifier exploits physically sensible information rather than behaving as an opaque black box.  
For the \texttt{XGB\_Combined} model, we therefore inspect the relative importance of each input feature using the standard XGBoost gain metric, and we further check the behaviour with SHapley Additive exPlanations (SHAP) values~\cite{lundberg2017unifiedapproachinterpretingmodel}; a detailed discussion of the interpretation is given in Appendix~\ref{app:XGB:importance_shap}.
Here we summarise the main points relevant for the physics interpretation.

\begin{figure*}[htbp]
  \centering
  \includegraphics[width=0.7\textwidth]{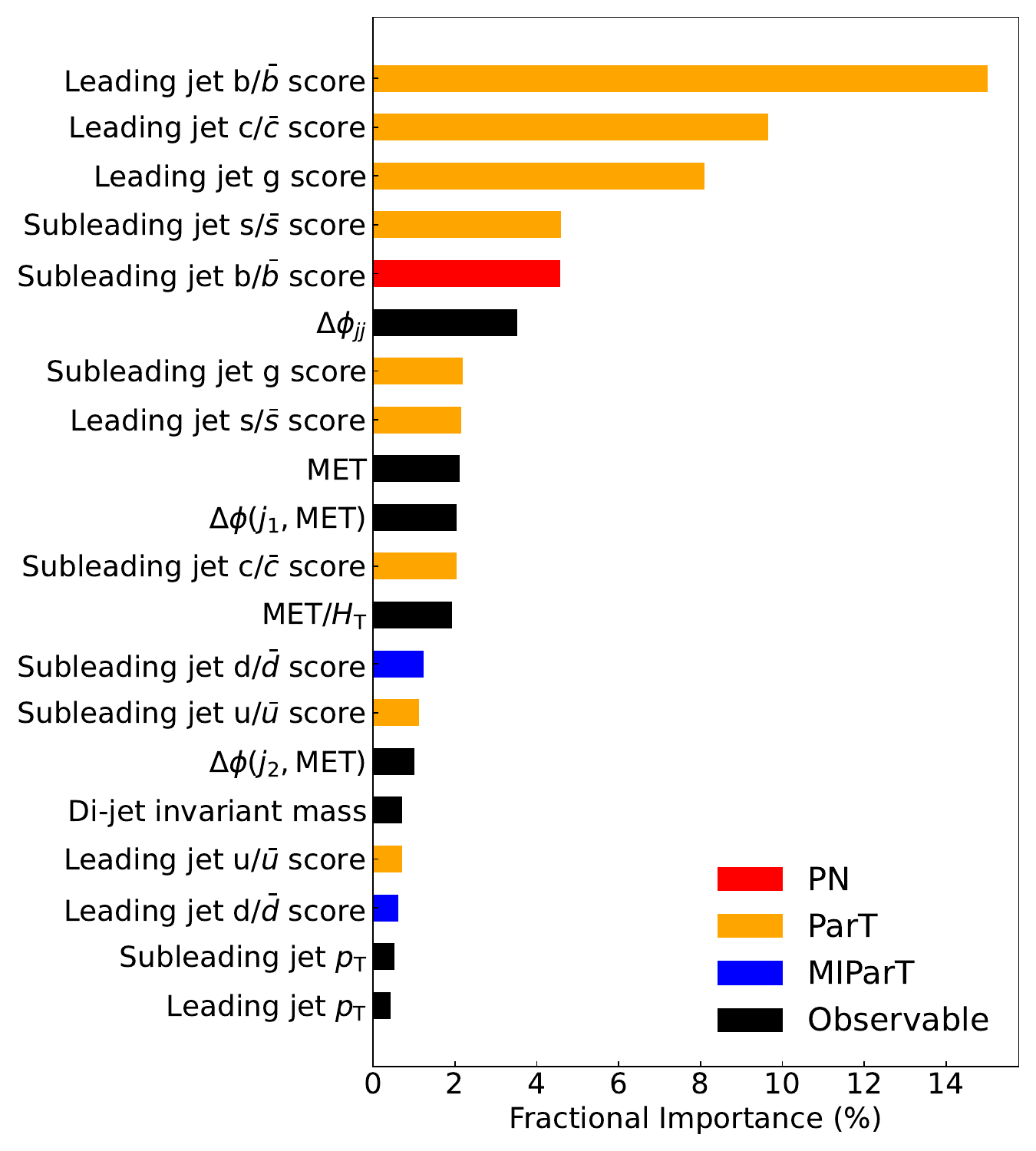}
  \caption{
    Top 20 most important input variables for the
    \texttt{XGB\_Combined} classifier, ranked by their total gain
    fraction in the XGBoost ensemble. Entries are
    colour-coded according to their origin: PN, ParT, MIParT,
    or high-level kinematic observables. 
  }
  \label{fig:xgb_top20_importance}
  \vspace{-0.2cm}
\end{figure*}

Fig.~\ref{fig:xgb_top20_importance} displays the twenty most important input variables for \texttt{XGB\_Combined}, ranked by their total gain fraction. The leading feature is the ParT $b$-jet score of the leading jet, followed by the $c$-jet and gluon scores. The next group consists of the ParT $s$-jet score and the PN $b$-jet score for the subleading jet.  This pattern confirms that heavy-flavor tagging provides the primary handle for isolating $H\to b\bar b$ and $H\to c\bar c$, while s- and g-jet information is essential for separating $H\to s\bar s$ and $H\to gg$.  The presence of PN and MIParT scores among the top features shows that the three jet taggers capture complementary aspects of the jet substructure, which the combined classifier can effectively exploit.  

The relative importance of the different taggers can be quantified by retraining the event-level classifier using only the ParT, MIParT or PN scores as jet inputs, while keeping the same set of kinematic observables.  As summarised in Table~\ref{tab:bdt_part_comparison} (Appendix~\ref{app:XGB:importance_shap}), using only the ParT scores already captures almost all of the tagging power.  Including the PN and MIParT scores in the \texttt{XGB\_Combined} setup leads to small but noticeable improvements, particularly for $H\to gg$ and $H\to s\bar s$, and reduces the spread among independent trainings (Fig.~\ref{fig:precision_comparison}), indicating that the multi-architecture ensemble is most beneficial in the difficult, light-flavor channels.

A complementary, event-by-event view is provided by the per-class SHAP analysis in Appendix~\ref{app:XGB:importance_shap}.  In brief, large $b$- or $c$-jet probabilities push events towards the $H\to b\bar b$ and $H\to c\bar c$ classes, while high strange or gluon scores shift events towards the $H\to s\bar s$ and $H\to gg$ hypotheses.  Missing-energy observables and $m_{jj}$ dominantly drive the separation between $ZH(\nu\bar\nu)$ and the $2f/4f$ backgrounds.  Taken together, these studies show that the classifier performance is rooted in well-understood heavy- and light-flavor tagging information combined with global event kinematics, rather than in unphysical artefacts of the training.

\subsection{Expected sensitivities and discussion}
\label{sec:results_sens}

We now quantify the sensitivity to hadronic Higgs decays by using the event-level classifier outputs as one-dimensional discriminants, as outlined in Sec.~\ref{sec:twostage}. For each Higgs decay hypothesis, the corresponding XGBoost score is used to select a signal-enhanced region in which all other Higgs modes and the $2f$/$4f$ processes are treated as backgrounds.  The threshold on the score is scanned, and for each working point the expected signal ($S$) and background ($B$) yields are computed. The statistical significance $Z$ is then evaluated using Eq.~\eqref{eq:significance}, and the optimal working point is chosen as the threshold that maximizes $Z$.

\begin{table*}[htbp]
\centering
\caption{Selection efficiencies [\%] for the four Higgs decay channels of interest. Each row shows the efficiency when the corresponding decay mode is treated as signal and all other Higgs modes and $2f/4f$ processes are treated as background.}
\label{tab:ana_eff}
\renewcommand{\arraystretch}{1.2}
\setlength{\tabcolsep}{3pt}
\vspace{0.2cm}
\small
\begin{tabular}{@{\hskip 0.1cm}cccccccc@{\hskip 0.1cm}}
\toprule
\multirow{2}{*}{\textbf{Signal}} & \multicolumn{7}{c}{\textbf{Channel}} \\
\cline{2-8}
& $H\to b\bar{b}$ & $H\to c\bar{c}$ & $H\to gg$ & $H\to s\bar{s}$ & $H\to ZZ^*$ & 4f & 2f \\
\midrule
$\boldsymbol{H\to b\bar{b}}$ & 64.81 & $3.31\times10^{-2}$ & 0.87 & $2.28\times10^{-2}$ & $1.33\times10^{-5}$ & $1.33\times10^{-1}$ & $6.84\times10^{-3}$ \\
$\boldsymbol{H\to c\bar{c}}$ & 0.03 & 52.56 & 0.58 & $3.65\times10^{-2}$ & $1.33\times10^{-5}$ & 0.10 & $2.20\times10^{-3}$ \\
$\boldsymbol{H\to gg}$ & 0.09 & 0.25 & 52.56 & 0.97 & 0 & $1.70\times10^{-2}$ & $1.26\times10^{-3}$ \\
$\boldsymbol{H\to s\bar{s}}$ & $6.00\times10^{-4}$ & 0.07 & 0.44 & 45.99 & 0 & 0.04 & $8.00\times10^{-4}$ \\
\bottomrule
\end{tabular}
\end{table*}

\vspace{-0.5cm}

\begin{table*}[htbp]
\centering
\caption{Event yields for the four Higgs decay channels of interest, normalized to $20~\text{ab}^{-1}$. Each row corresponds to the case where that decay mode is treated as signal and all other modes are treated as background.}
\label{tab:ana_yields}
\renewcommand{\arraystretch}{1.2}
\setlength{\tabcolsep}{7pt}
\vspace{0.2cm}
\small
\begin{tabular}{@{\hskip 0.4cm}cccccccc@{\hskip 0.4cm}}
\toprule
\multirow{2}{*}{\textbf{Signal}} & \multicolumn{7}{c}{\textbf{Channel}} \\
\cline{2-8}
& $H\to b\bar{b}$ & $H\to c\bar{c}$ & $H\to gg$ & $H\to s\bar{s}$ & $H\to ZZ^*$ & 4f & 2f \\
\midrule
$\boldsymbol{H\to b\bar{b}}$ & 346212.75 & 8.95 & 687.06 & 0.05 & $9.07\times10^{-4}$ & 9868.37 & 1805.06 \\
$\boldsymbol{H\to c\bar{c}}$ & 170.60 & 14191.65 & 462.57 & 0.07 & $9.07\times10^{-4}$ & 7139.01 & 580.58 \\
$\boldsymbol{H\to gg}$ & 484.72 & 67.52 & 41734.45 & 1.93 & 0 & 1256.07 & 332.51 \\
$\boldsymbol{H\to s\bar{s}}$ & 3.21 & 20.05 & 346.85 & 91.97 & 0 & 3288.52 & 211.25 \\
\bottomrule
\end{tabular}
\end{table*}

Table~\ref{tab:ana_eff} and Table~\ref{tab:ana_yields} summarize the resulting selection efficiencies and event yields for the four Higgs decay channels of interest, using the \texttt{XGB\_Combined} classifier and assuming $20~\text{ab}^{-1}$.
Using the sample sizes in Table~\ref{tab:sim_samples} and the corresponding selection efficiencies, we made a simple estimate of the Monte Carlo statistical uncertainty from the finite selected-event counts for each background component. The resulting fluctuation on the total background yield is small for all four signal regions.
For illustration, when targeting $H\to s\bar s$ we obtain about 92 signal events with a background of about 3800 events, corresponding to a statistical significance of $Z\simeq1.5\sigma$.  Although this does not yet constitute an observation, it nevertheless provides a sensitivity estimate for $H\to s\bar s$ using a single channel $e^+e^- \to ZH \to\nu\bar{\nu}s\bar s$ at a lepton collider and provides a concrete benchmark for future improvements which will include more $Z$ decay channels (e.g., $Z\to\ell^+\ell^-, q\bar q$).

\begin{figure*}[htbp]
\vspace{-0.5cm}
    \centering
    \includegraphics[width=0.9\linewidth]{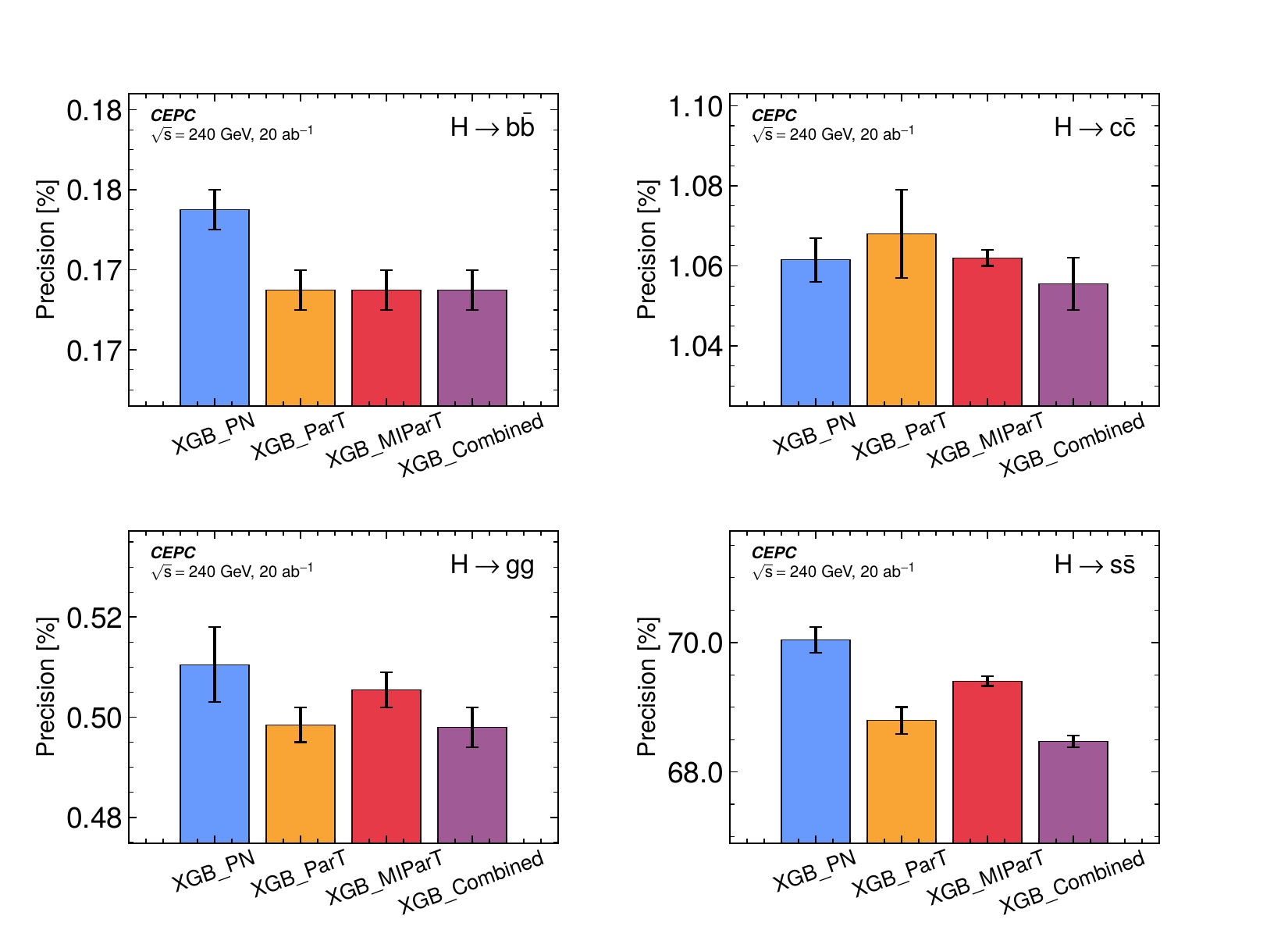}
    \caption{
        Relative precision on different Higgs decay channels for the four
        event-level classifiers: \texttt{XGB\_PN}, \texttt{XGB\_ParT},
        \texttt{XGB\_MIParT}, and \texttt{XGB\_Combined}.  For each classifier, 5 independent training runs are performed; the bars show the mean precision and the error bars indicate the full variation range (max–min) among the runs.  
        Panels correspond to $H\to b\bar{b}$, $H\to c\bar{c}$, $H\to gg$, and $H\to s\bar{s}$, respectively.
    }
    \label{fig:precision_comparison}
    \vspace{-0.5cm}
\end{figure*}

To assess the stability of the results and the benefit of combining different jet taggers, we repeat the full analysis chain for each of the four event-level classifiers: \texttt{XGB\_PN}, \texttt{XGB\_ParT}, \texttt{XGB\_MIParT} and \texttt{XGB\_Combined}.  For every classifier, five statistically independent training runs are performed, yielding a distribution of projected precisions for each Higgs decay channel.  The results are summarized in Fig.~\ref{fig:precision_comparison}, where the bars indicate the mean precision and the error bars span the full range (maximum–minimum) observed across the five runs.

For $H\to b\bar b$ and $H\to c\bar c$, all four classifiers achieve similar sub-permille and percent-level precisions, respectively.  The combination of the three jet taggers brings only marginal improvement, which is expected given the already high purity and strong kinematic separation of these modes.  For $H\to gg$, the combined classifier provides a noticeable gain: the central precision improves and, more importantly, the spread among independent trainings is reduced, indicating that the ensemble is more robust against statistical fluctuations in the training process.  In the most challenging $H\to s\bar s$ channel, the projected precision remains at the level of $\sim 68\%$, reflecting the limited signal statistics and the residual overlap with $H\to gg$ and $4f$ backgrounds. 
Nevertheless, the \texttt{XGB\_Combined} classifier yields the best central value, and its entire variation band lies below those of the other three models, demonstrating that the multi-architecture combination is beneficial precisely where the problem is most difficult.

\begin{table*}[htbp]
    \centering
    \caption{
        Comparison of relative measurement precision for various Higgs decay channels between the CEPC Higgs benchmark ~\cite{An_2019}, FCC-ee~\cite{DelVecchio:2025gzw} and this work, normalized to $20~\text{ab}^{-1}$.  The last column shows the relative improvement in precision with respect to the CEPC benchmark.
    }
    \label{tab:precision_comparison}
    \setlength{\tabcolsep}{10pt}
    \begin{tabular}{@{\hskip 0.5cm}c c c c c@{\hskip 0.5cm}}
        \toprule
        \textbf{Channel} & \textbf{CEPC} & \textbf{FCC-ee} &\textbf{This work} & \textbf{Imp. vs CEPC} \\
        \midrule
        $Z \to \nu\bar{\nu},~ H \to b\bar{b}$ &0.20\%       &  0.28\%& 0.17\% & 15\% \\
        $Z \to \nu\bar{\nu},~ H \to c\bar{c}$ &1.85\%       &  1.95\%& 1.06\% & 43\% \\
        $Z \to \nu\bar{\nu},~ H \to gg$       &0.70\%       &  0.78\%& 0.50\% & 29\% \\
        $Z \to \nu\bar{\nu},~ H \to s\bar{s}$ &  ---        &   117\%   & 68\% & --- \\
        \bottomrule
    \end{tabular}
    \vspace{-0.3em}
\end{table*}

A direct comparison with previous CEPC and FCC-ee projections is given in Table~\ref{tab:precision_comparison}.  
The ``CEPC'' column quotes the relative precisions on $\sigma(ZH)\times\text{Br}(H\to X)$ obtained in the CEPC Higgs benchmark study~\cite{An_2019} for the $\nu\bar\nu H$ channel, rescaled to $20~\text{ab}^{-1}$, while the ``FCC-ee'' column shows the corresponding projections from the FCC-ee study~\cite{DelVecchio:2025gzw}. 
The ``This work'' column gives the results from the \texttt{XGB\_Combined} classifier in this study.
For $H\to c\bar c$ the precision improves from $1.85\%$ to $1.06\%$, corresponding to a relative gain of about $43\%$.  For $H\to gg$ the precision improves from $0.70\%$ to $0.50\%$ (about $29\%$ gain), and for $H\to b\bar b$ a modest but non-negligible improvement from $0.20\%$ to $0.17\%$ is observed.  The $H\to s\bar s$ mode was not included in the CEPC benchmark study; our study provides a quantitative estimate, with a projected precision of about $68\%$. 
Relative to the FCC-ee projections, the uncertainties are reduced by about $39\%$ and $46\%$ for $H\to b\bar b$ and $H\to c\bar c$, respectively, and by about $36\%$ and $42\%$ for $H\to gg$ and $H\to s\bar s$, respectively. 
These quantitative differences with respect to previous CEPC and FCC-ee studies should be interpreted with care, as the underlying assumptions are not fully identical. They may reflect differences in detector performance, event reconstruction, generator setup, multivariate strategy, and the treatment of systematic uncertainties. In particular, the present results are statistical-only projections.

The improvements for $H\to c\bar c$ and $H\to gg$ demonstrate the impact of modern particle-level deep learning on Higgs measurements at future $e^+e^-$ colliders.
By directly exploiting the fine-grained jet substructure and combining multiple complementary taggers in a two-stage analysis, we achieve substantially better separation between quark and gluon jets and among different quark flavors than methods based solely on hand-crafted observables.

The $H\to s\bar s$ channel remains statistically limited in the baseline CEPC configuration considered here.  However, the present analysis shows that even with $20~\text{ab}^{-1}$ it is possible to reach a sensitivity at the level of $Z\sim1.5\sigma$.  Future lepton colliders with higher luminosity, or combined analyses including additional $Z$ decay modes beyond $Z\to\nu\bar\nu$, could further enhance the sensitivity and discovery potential.  
In addition, systematic uncertainties, such as those associated with jet energy scale, flavor-tagging calibration, and theoretical modeling of $2f/4f$ backgrounds, have not been included in the present study and will need to be assessed in more realistic scenarios.  Nevertheless, the framework developed here is readily extendable: once experimental calibrations are available, the same two-stage strategy can incorporate data-driven constraints and systematic nuisance parameters in a global fit.

The sensitivities quoted above are baseline statistical projections obtained under the detector assumptions implemented in \textsc{Delphes}. In particular, the PID-related inputs used in the jet-tagging networks are taken from the parameterized CEPC detector response and therefore do not fully capture detector-level fake rates or PID calibration uncertainties. To estimate the impact of plausible detector-related systematic effects, we consider two variations. First, guided by the CEPC Reference TDR, moving from an idealized PID scenario to a more realistic one is expected to reduce the single $s$-jet tagging efficiency from about 70\% to about 60\%. Since the $H\to s\bar{s}$ analysis relies on double $s$-tagging, this corresponds to an estimated reduction of the $H\to s\bar{s}$ signal yield by about 27\%, and, under the simplifying assumption of unchanged background yield, a degradation of the expected significance from about $1.5\sigma$ to about $1.1\sigma$. Second, we apply an additional $\sim 5\%$ smearing~\cite{CEPCStudyGroup:2025kmw} to the jet kinematics to mimic degraded jet-energy resolution and repeat the full analysis chain. The resulting yield variations are modest overall, and the main qualitative conclusions remain unchanged. These studies provide a first assessment of the robustness of the analysis against plausible detector effects. We next compare the present two-stage framework with an alternative holistic event-classification strategy explored in previous work.

Previous work has explored an alternative ``holistic'' classification strategy~\cite{Zhu:2025eoe}, in which the full reconstructed event is fed to an event classifier realized using ParticleNet. 
When the comparison is restricted to the $\nu\nu H$ channel with $H\to b\bar{b},\,c\bar{c},\,gg,\,s\bar{s}$ as signal and the $2f$ and $4f$ backgrounds are omitted, the resulting statistical precisions are very similar 
to those obtained with our two-stage framework, providing a useful handle for validating jet-origin identification. 
This indicates that the two-stage design retains essentially the same statistical discrimination power in this simplified setup, while offering a more modular and physically interpretable analysis procedure.
A detailed numerical comparison and discussion are given in Appendix~\ref{app:holistic}. They further observed that using \textsc{Pythia}~8.313 instead of \textsc{Pythia}~6.4 improves the relative precision of the $H\to s\bar{s}$ final state by about 40\%, in particular by producing a higher rate of $K^{\pm}$ mesons, which in turn enhances the separation of s-jets. 

\section{Conclusions}
\label{sec:conclusion}

In this paper we have studied the prospects for measuring hadronic Higgs decays at a future high-luminosity $e^+e^-$ Higgs factory, using the CEPC as a concrete benchmark. Focusing on $e^+e^-\to ZH$ with $Z\to\nu\bar\nu$ and $H\to q\bar q,gg$, we developed a unified two-stage analysis framework that combines three state-of-the-art particle-level jet taggers, ParticleNet, Particle Transformer (ParT) and More-Interaction Particle Transformer (MIParT), with an event-level XGBoost classifier. This approach efficiently exploits both jet substructure and global event kinematics to separate $H\to b\bar b$, $H\to c\bar c$, $H\to s\bar s$ and $H\to gg$ from the dominant $2f$ and $4f$ backgrounds.

Assuming an integrated luminosity of $20~\text{ab}^{-1}$ at $\sqrt{s}=240~\text{GeV}$, we obtain projected relative uncertainties on $\sigma(ZH)\times\text{Br}(H\to X)$ of $0.17\%$ for $X=b\bar b$, $1.06\%$ for $c\bar c$, $0.50\%$ for $gg$ and $68\%$ for $s\bar s$ in the $Z\to\nu\bar\nu$ channel. Compared with the CEPC Higgs benchmark study, the precisions for $H\to c\bar c$ and $H\to gg$ are improved by about $43\%$ and $29\%$, respectively, while $H\to b\bar b$ improves modestly from $0.20\%$ to $0.17\%$. For $H\to s\bar s$ we provide a quantitative sensitivity estimate at a lepton collider, corresponding to a statistical significance of about $1.5\sigma$ and establishing a concrete benchmark for future probes of the strange-quark Yukawa coupling.

Although the present study considers only statistical uncertainties and a single $Z$ decay mode, it already illustrates the potential of particle-level deep learning for Higgs physics at lepton colliders.  A more complete assessment will require the inclusion of experimental and theoretical systematics, detailed detector calibrations (in particular for jet energy scale and flavor tagging), and the combination of multiple $ZH$ final states such as $Z\to\ell^+\ell^-$ and $Z\to q\bar q$.  Higher integrated luminosities, or staged running at different center-of-mass energies, would further enhance the reach for rare decays such as $H\to s\bar s$.

The sensitivity of $H\to ss$ measurements can be enhanced via three key avenues: improved detector performance, advanced jet origin identification algorithms, and rigorously validated hadronization models. At the current analysis threshold, the dominant backgrounds to $H\to ss$ are Standard Model four-fermion processes with identical parton color configurations and $H\to gg$ processes. The former background can be suppressed by optimizing the reconstruction of the total invariant mass and recoil mass of the hadronic system—an objective aligned with the emerging concept of one-to-one correspondence detectors. Reducing contamination from $H\to gg$ processes, by contrast, relies on enhanced discrimination between gluon-initiated and strange-quark-initiated jets; as demonstrated in Ref.~\cite{Zhu:2025eoe}, accurate modeling and validation of hadronization frameworks are critical to this end. It is also worth noting that $H\to ss$ measurements remain hampered by inherent theoretical challenges~\cite{Altmann:2025feg}, which necessitate targeted follow-up studies. 

Finally, while our numerical results are obtained for the CEPC reference design, the analysis strategy developed here is readily applicable to other proposed Higgs factories, including FCC-ee and ILC, by updating the detector response and luminosity assumptions.  Together with the gains demonstrated for $H\to c\bar c$ and $H\to gg$ and a quantitative projection for $H\to s\bar s$, this work highlights the important role that advanced machine-learning techniques can play in fully mapping out the flavor structure of the Higgs sector at future lepton colliders.

\acknowledgments
We thank Chongqiao Li and Kaili Zhang for useful discussions and comments. This work was supported by the National Natural Science Foundation of China under Grant No. W2441004. 
The work of Kun Wang was supported by the Open Project of the Shanghai Key Laboratory for Particle Physics and Cosmology under Grant No. 22DZ2229013-3.

\appendix
\section{Preselection kinematics}
\label{app:preselection}

\begin{figure*}[htbp]
{
    \centering

    \begin{tabular}{cc}
        \includegraphics[width=0.49\linewidth]{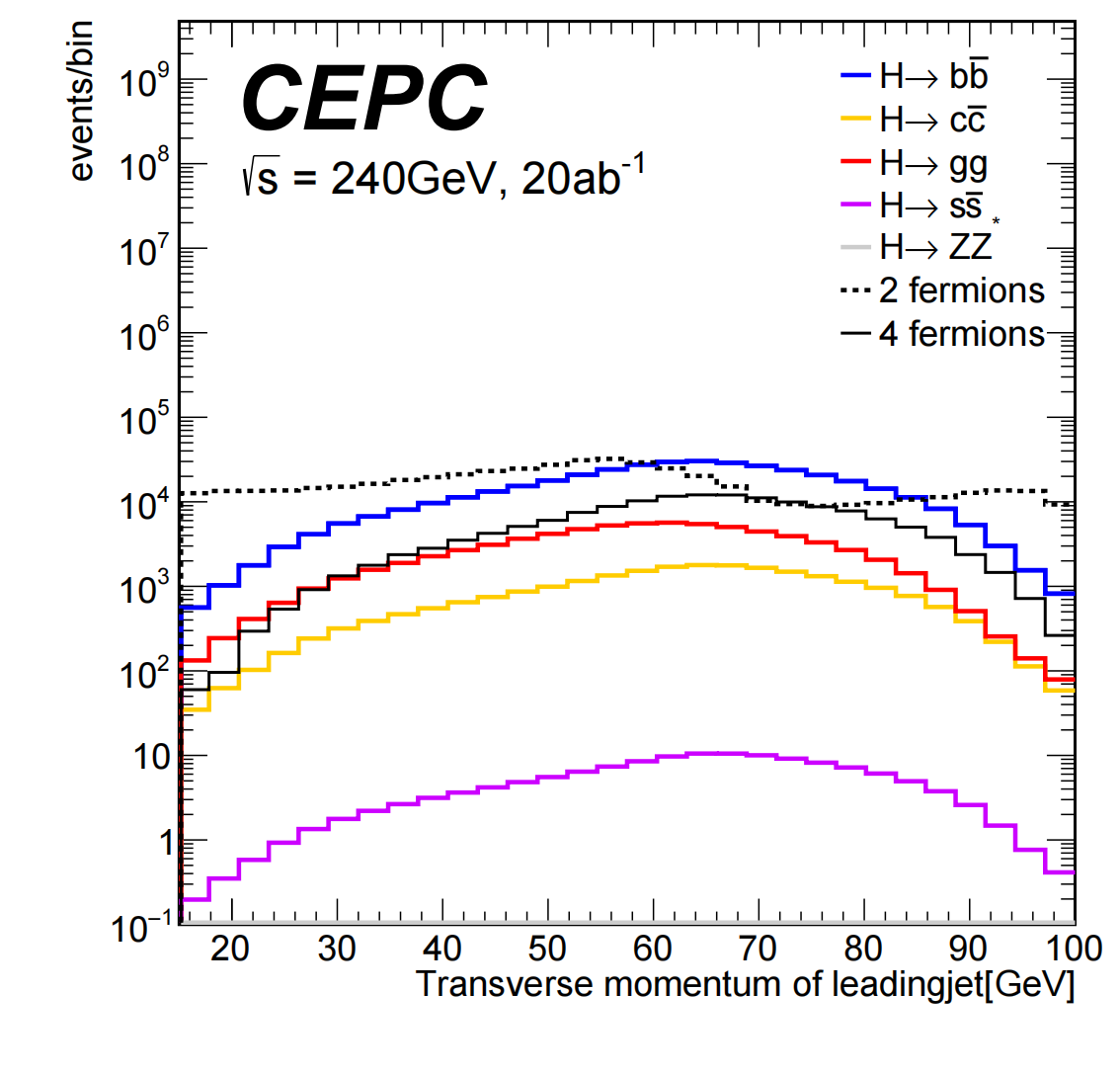} &
        \includegraphics[width=0.49\linewidth]{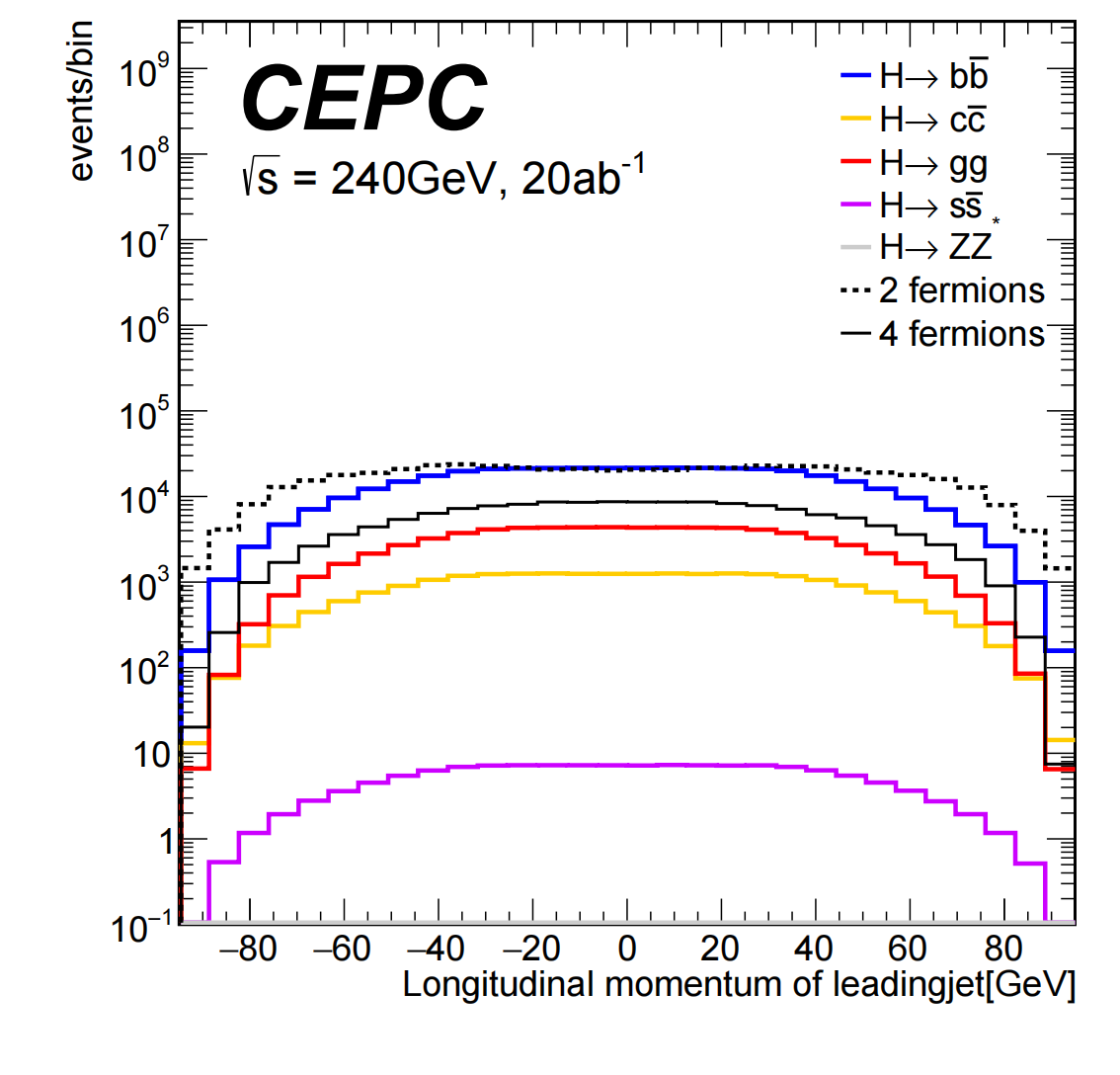} \\
        \textbf{(a)} & \textbf{(b)} \\
    \end{tabular}

    \vspace{0.6em}

    \begin{tabular}{cc}
        \includegraphics[width=0.49\linewidth]{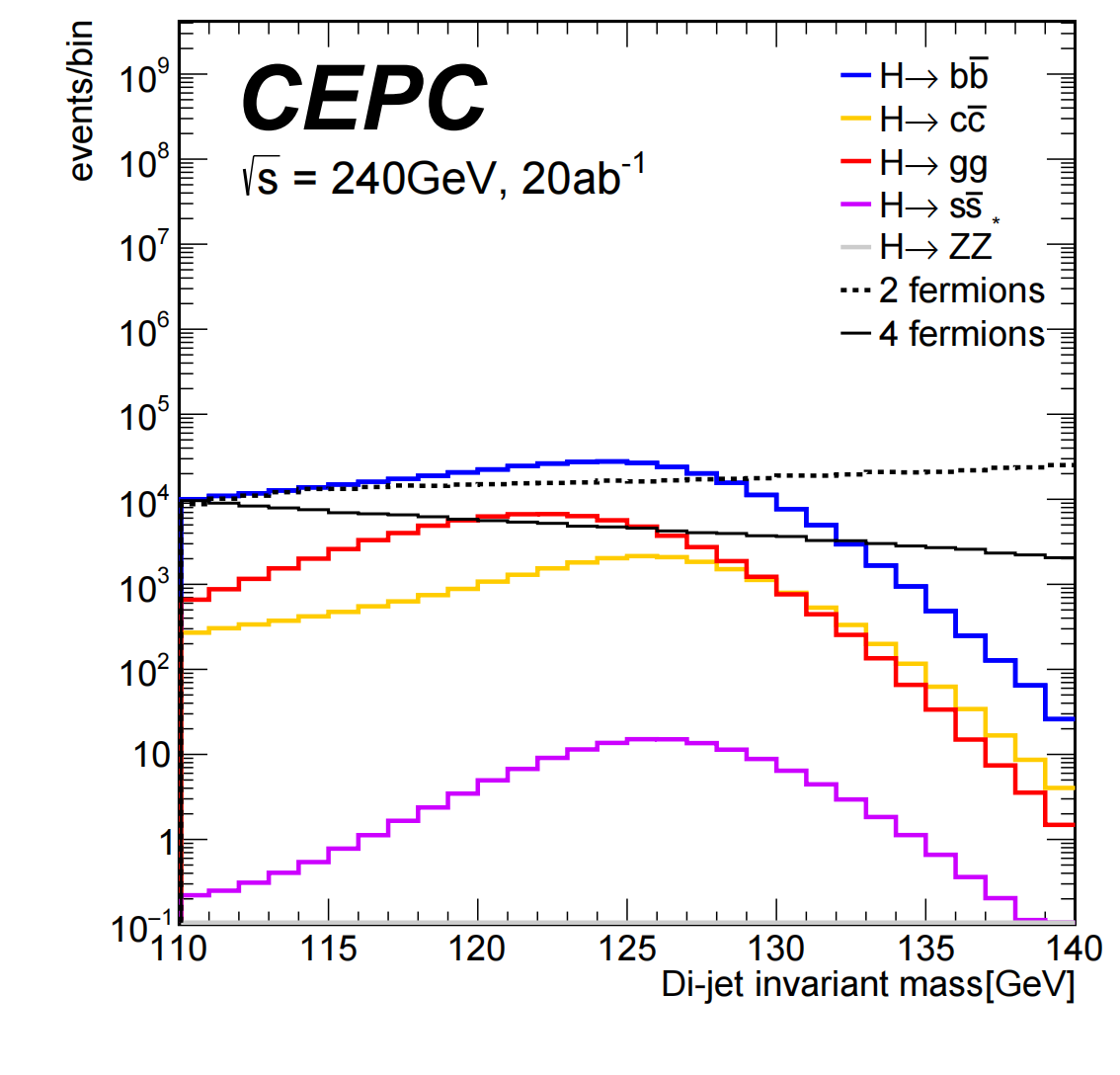} &
        \includegraphics[width=0.49\linewidth]{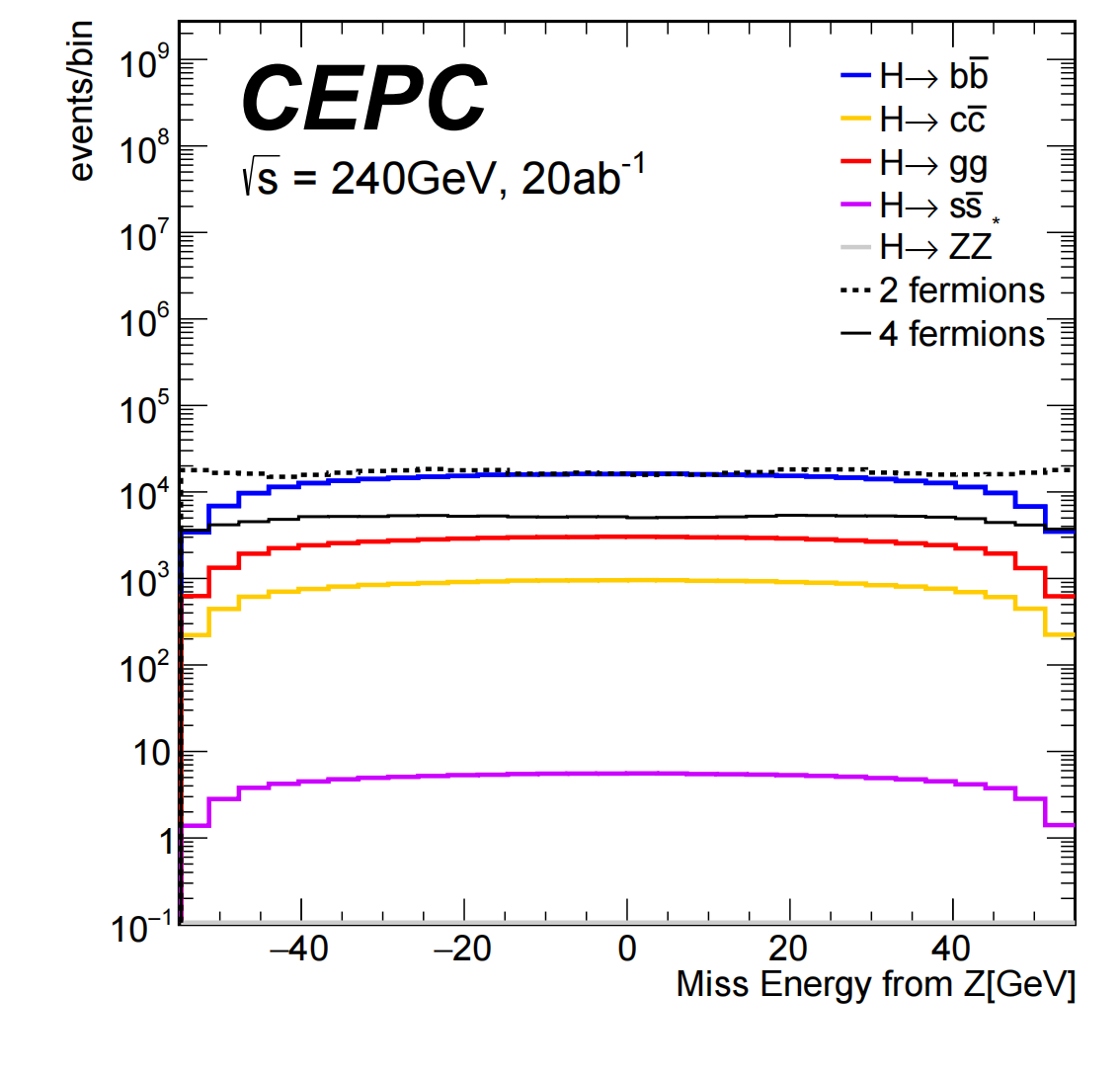} \\
        \textbf{(c)} & \textbf{(d)} \\
    \end{tabular}
}

\caption{
    Distributions of the preselection observables used as inputs to the
    event-level multivariate analysis:
    (a)~transverse momentum $p_\textup{T}(j)$ and
    (b)~longitudinal momentum $p_{z}(j)$ of the leading jet,
    (c)~invariant mass $m_{jj}$ of the dijet system,
    and (d)~longitudinal component of the missing energy $E_{z,\text{miss}}$.
    All distributions are shown after event-quality selections and
    normalised to $20~\text{ab}^{-1}$.
}
\label{fig:preselection_vars}
\end{figure*}

As discussed in Sec.~\ref{sec:twostage}, a loose preselection is applied before training the event-level XGBoost classifiers.   The aim of this step is twofold:   (i) to enhance the $ZH(\nu\bar{\nu})$ topology and suppress the large inclusive $2f$ and $4f$ backgrounds, and   (ii) to stabilise the kinematic phase space over which the multivariate classifier is trained, so that it can focus on finer differences related to jet flavour. The preselection uses four global observables: the transverse ($p_\textup{T}$) and longitudinal ($p_z$) momenta of the leading jet, the invariant mass of the dijet system $m_{jj}$, and the longitudinal component of the missing energy $E_{z,\text{miss}}$. Figures~\ref{fig:preselection_vars}(a)--(d) show the distributions of these variables for the main signal and background processes, after basic event-quality requirements and normalised to an integrated luminosity of $20~\text{ab}^{-1}$.

The leading-jet transverse momentum $p_\textup{T}(j)$, shown in Fig.~\ref{fig:preselection_vars}(a), reflects the recoil of the Higgs boson against the $Z\to\nu\bar{\nu}$ system. In $e^+e^-\to ZH$ at $\sqrt{s}=240~\text{GeV}$ the Higgs is produced close to threshold, so that its decay products carry typical transverse momenta of order $m_H/2$. Consequently, all hadronic Higgs decay modes peak in a relatively narrow range around $p_\textup{T}(j)\sim 50$--$70~\text{GeV}$. In contrast, the $2f$ background $e^+e^-\to q\bar q$ originates from $\gamma^*/Z$ exchange and exhibits a much broader spectrum, extending to both lower and higher $p_\textup{T}$ values, while the $4f$ processes populate mainly the lower-$p_\textup{T}$ region. Requiring $15 < p_\textup{T}(j) < 100~\text{GeV}$ therefore retains the bulk of the Higgs signal while efficiently rejecting very soft and very hard $2f/4f$ topologies that are kinematically incompatible with $ZH$.

The longitudinal momentum $p_z(j)$ of the leading jet, displayed in Fig.~\ref{fig:preselection_vars}(b), encodes the polar-angle distribution of the event. Since the Higgs boson is predominantly produced centrally in $e^+e^-\to ZH$, the jets from its decay tend to be central as well, resulting in a fairly flat distribution in $p_z$ within a limited range. By contrast, the $2f$ background is strongly peaked at forward angles, leading to an enhanced rate at large $|p_z(j)|$. The symmetric requirement $|p_z(j)|<95~\text{GeV}$ exploits this difference, suppressing a large fraction of the forward-peaked $2f$ events while leaving the $ZH$ signal essentially unaffected.

Fig.~\ref{fig:preselection_vars}(c) shows the invariant mass of the dijet system, $m_{jj}$. For genuine $ZH$ events the two reconstructed jets originate from the Higgs decay and therefore cluster around the Higgs boson mass, with a resolution of a few~GeV determined by the detector and reconstruction performance. This leads to the characteristic peak near $m_{jj}\simeq 125~\text{GeV}$ for all hadronic Higgs decay modes. In contrast, the $2f$ background forms a broad continuum in $m_{jj}$, while the $4f$ processes (including $\nu\bar{\nu}q\bar{q}$) exhibit a wider distribution reflecting different intermediate boson masses and off-shell configurations. The mass window $110 < m_{jj} < 140~\text{GeV}$ is therefore an efficient handle to enhance the Higgs contribution and is also used as an input feature for the multivariate analysis.

Finally, Fig.~\ref{fig:preselection_vars}(d) presents the distribution of the longitudinal component of the missing energy, $E_{z,\text{miss}}$, which is reconstructed from the vector sum of all visible final-state particles. In signal events, the missing momentum is dominated by the $Z\to\nu\bar{\nu}$ decay and is thus correlated with the recoil of the Higgs system, leading to a relatively broad distribution in $E_{z,\text{miss}}$ that is symmetric around zero. For the $2f$ background there is no genuine source of missing energy; the small apparent $E_{z,\text{miss}}$ arises from detector resolution and acceptance losses and is therefore concentrated near zero. The $4f$ background lies in between these two extremes. Imposing $|E_{z,\text{miss}}|<55~\text{GeV}$ suppresses extreme configurations in which the apparent missing momentum is aligned with the beam direction, which are predominantly due to background or poorly reconstructed events.

Taken together, these four observables capture the essential kinematic features of the $ZH(\nu\bar{\nu})$ topology: a centrally produced Higgs boson decaying to two jets with an invariant mass near $m_H$ and recoiling against genuine missing energy. The preselection cuts quoted in Sec.~\ref{sec:twostage} retain about $80\%$ of the Higgs signal while rejecting more than $97\%$ of the inclusive $2f$ and $4f$ backgrounds. At the same time, the full distributions shown in Fig.~\ref{fig:preselection_vars} are used as inputs to the XGBoost classifier, allowing the multivariate analysis to exploit the residual shape differences beyond the simple rectangular cuts.

\section{Event-level classifier details}
\label{app:xgb_details}

In this appendix we provide additional information on the training of the XGBoost event-level classifiers and discuss, from a physics perspective, which input variables drive the separation between signal and background hypotheses.

\subsection{XGBoost setup and training strategy}
\label{app:XGB:hyper}

The event classification is performed with gradient-boosted decision trees as implemented in \textsc{XGBoost}~\cite{XGBoost}.  We employ a multi-class setup in which six classes are distinguished simultaneously: $H\to b\bar b$, $H\to c\bar c$, $H\to s\bar s$, $H\to gg$, and the $2f$ and $4f$ backgrounds.  The classifier output is therefore a six-component probability vector for each event, and the score associated with a particular signal hypothesis is used as the one-dimensional discriminant in Sec.~\ref{sec:results_sens}.

The main hyperparameters are chosen so as to balance model complexity, training stability and robustness against overfitting:

\begin{itemize}
  \item The maximum depth of each decision tree is set to
        \texttt{max\_depth} $=3$.  

  \item The learning rate is scheduled in three stages:
        $\eta=0.1$ for the first 600 boosting rounds,
        $\eta=0.05$ for rounds 601--800, and
        $\eta=0.01$ for the last 200 rounds,
        with a total of \texttt{n\_estimators} $=1000$ trees.

  \item To introduce additional regularisation and decorrelation among
        trees we use row and column subsampling,
        \texttt{subsample} $=0.8$ and
        \texttt{colsample\_bytree} $=0.8$, respectively.

  \item The objective function is the multiclass logarithmic loss
        (\texttt{multi:softprob} in \textsc{XGBoost}),
        and the default $\ell_2$ regularisation parameters are retained
        as further stabilisers.
\end{itemize}

\subsection{Physics interpretation}
\label{app:XGB:importance_shap}

The XGBoost classifiers take as input both global event observables and the flavor probabilities predicted by the three jet taggers (PN, ParT and MIParT) for the two Higgs-candidate jets (cf.\ Table~\ref{tab:BDT_inputs}). To understand which information the classifier exploits most strongly, we study the relative importance of each input feature using two complementary measures:

\begin{enumerate}
  \item The total gain attributed to a feature, summed over all decision
        trees in the ensemble, as feature importance, which quantifies how much this feature
        contributes to reducing the training loss.

  \item SHapley Additive exPlanations (SHAP) values~\cite{lundberg2017unifiedapproachinterpretingmodel},
        which characterise, event by event, how much each input variable
        pushes the classifier output towards or away from a specific class.
        In the multi-class setting used here, a separate SHAP value is
        computed for each class.
\end{enumerate}

\subsubsection*{Global feature importance from gain}

Fig.~\ref{fig:xgb_top20_importance} shows the 20 most important input features for the \texttt{XGB\_Combined} classifier, ranked according to their total gain.
Each entry is colour-coded according to its origin: PN, ParT, MIParT or a high-level observable. Several physics-driven patterns are immediately visible.

We can see that the jet flavor scores are clearly dominant in the ranking.
In particular, the ParT scores for the leading-jet $b$ and $c$ hypotheses are the two most important features.  This is physically expected: heavy-flavor tagging provides the primary handle to separate $H\to b\bar b$ and $H\to c\bar c$ from the light-flavor Higgs decays and from the $2f/4f$ backgrounds. The corresponding scores for the subleading jet, as well as the strange and gluon probabilities for both jets, also appear high in the list. They are essential for distinguishing $H\to s\bar s$ and $H\to gg$ from each other and from the inclusive light-quark background. The fact that PN and MIParT scores for the same flavor categories also enter among the top-ranked variables demonstrates that the three jet taggers provide partially complementary information: XGBoost learns to assign different weights to their outputs depending on the flavor and on whether the jet is leading or subleading in $p_\textup{T}$.

Among the non-jet-flavor observables, the azimuthal separation $\Delta\phi_{jj}$ between the two jets is the most important. It is sensitive to differences in event shapes between $ZH$ production and generic $2f/4f$ processes.
Other kinematic variables like missing transverse energy also contribute. This reflects the role of the $Z\to\nu\bar\nu$ topology in separating signal from the $2f$ background, which has no genuine neutrinos and thus only small, resolution-driven missing energy.

\subsubsection*{Per-class SHAP analysis}
\begin{figure*}[!p]
{
    \centering

    \begin{tabular}{cc}
        \includegraphics[width=0.49\linewidth]{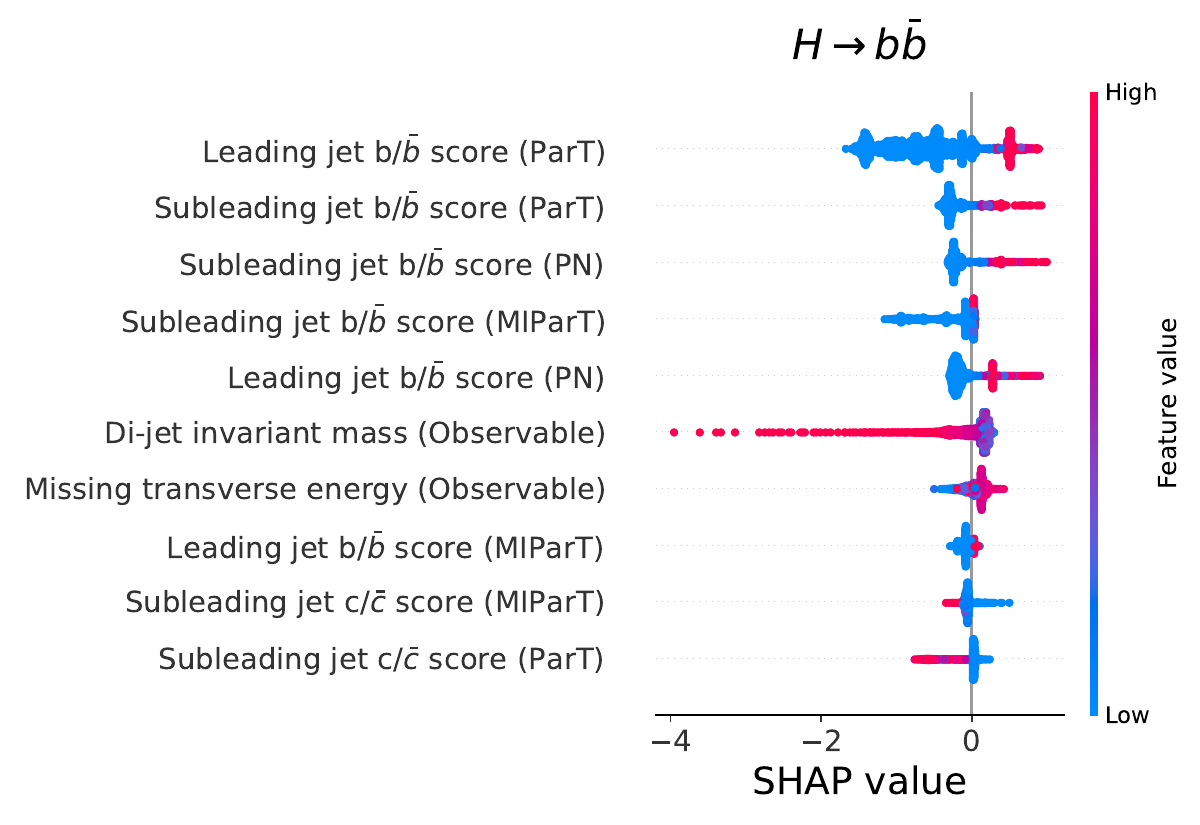} &
        \includegraphics[width=0.49\linewidth]{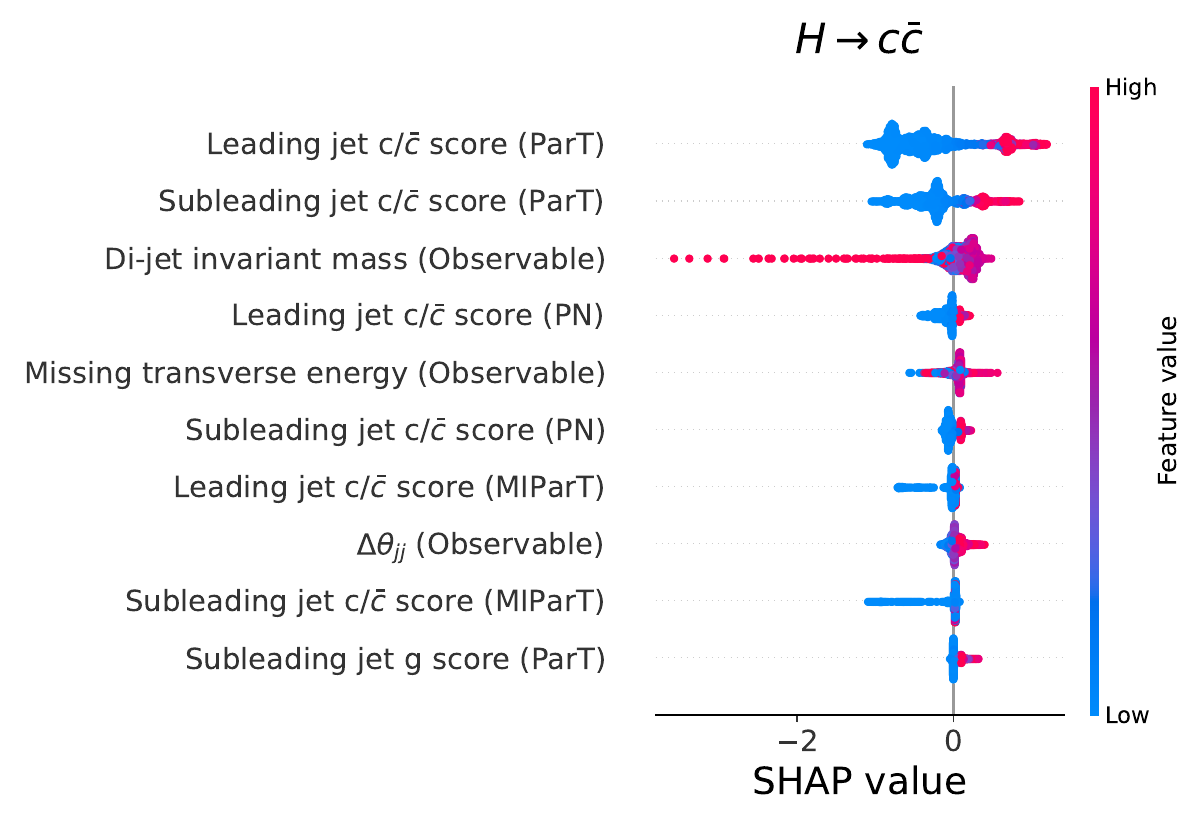} \\
        \textbf{(a)} & \textbf{(b)} \\
    \end{tabular}

    \vspace{0.8em}

    \begin{tabular}{cc}
        \includegraphics[width=0.49\linewidth]{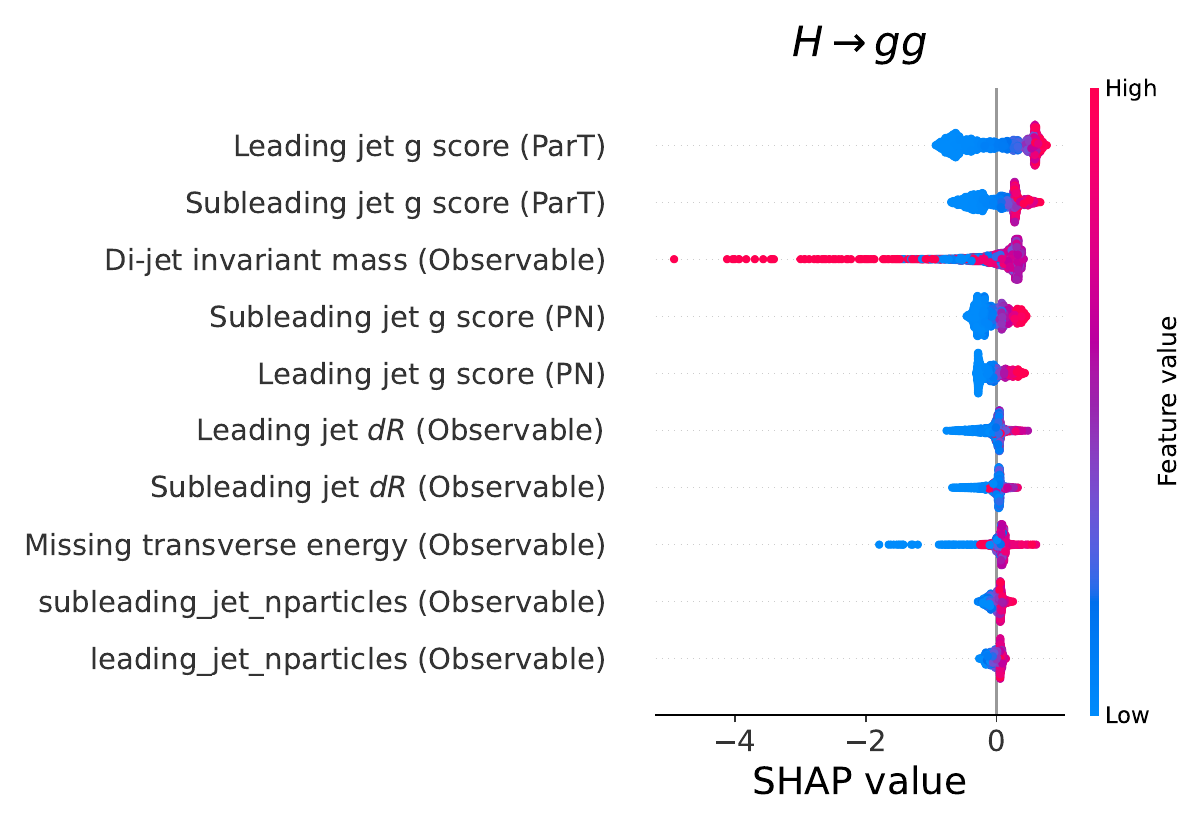} &
        \includegraphics[width=0.49\linewidth]{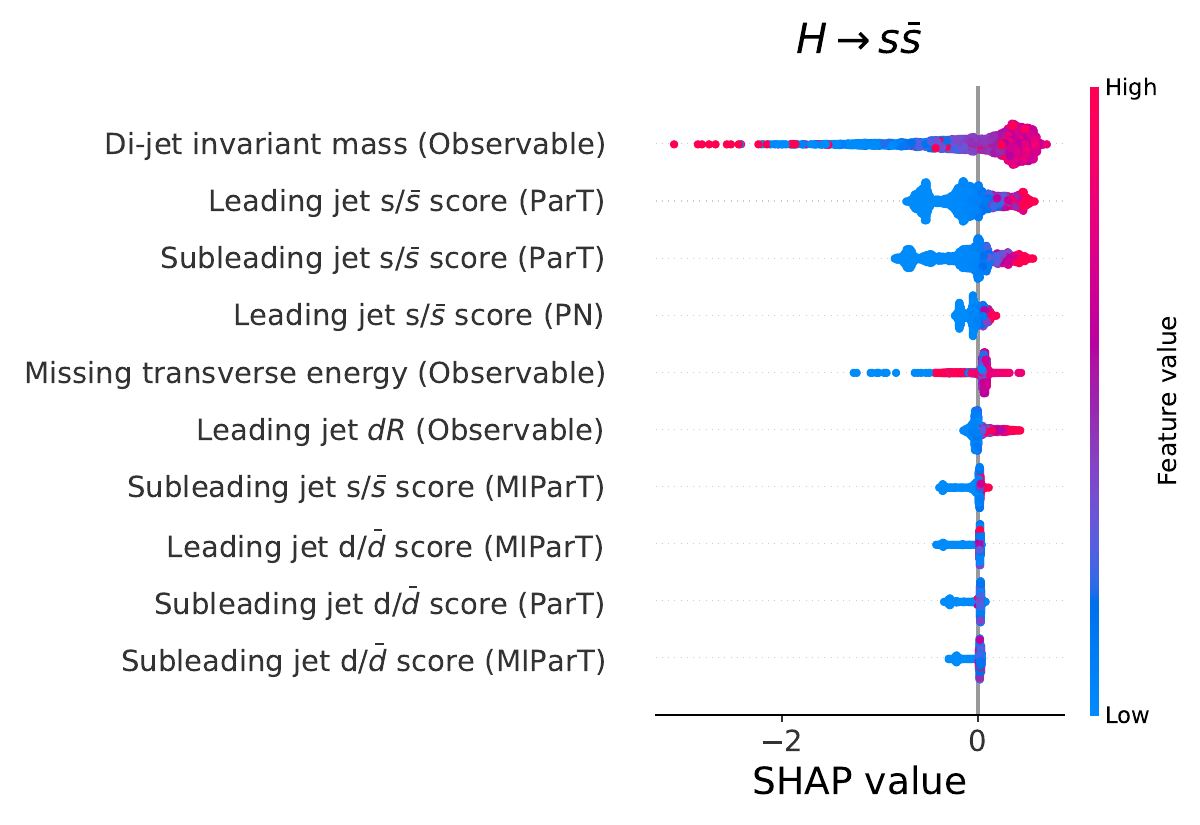} \\
        \textbf{(c)} & \textbf{(d)} \\
    \end{tabular}

    \vspace{0.8em}

    \begin{tabular}{cc}
        \includegraphics[width=0.49\linewidth]{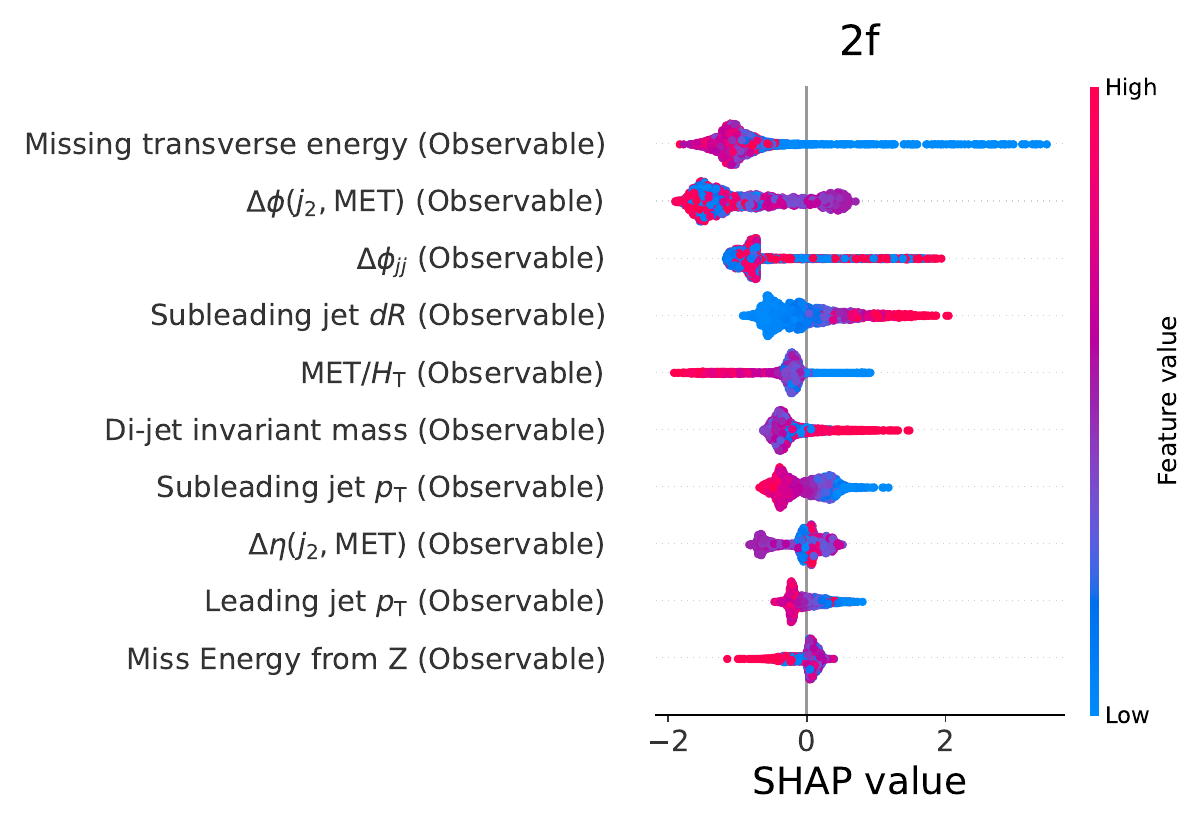} &
        \includegraphics[width=0.49\linewidth]{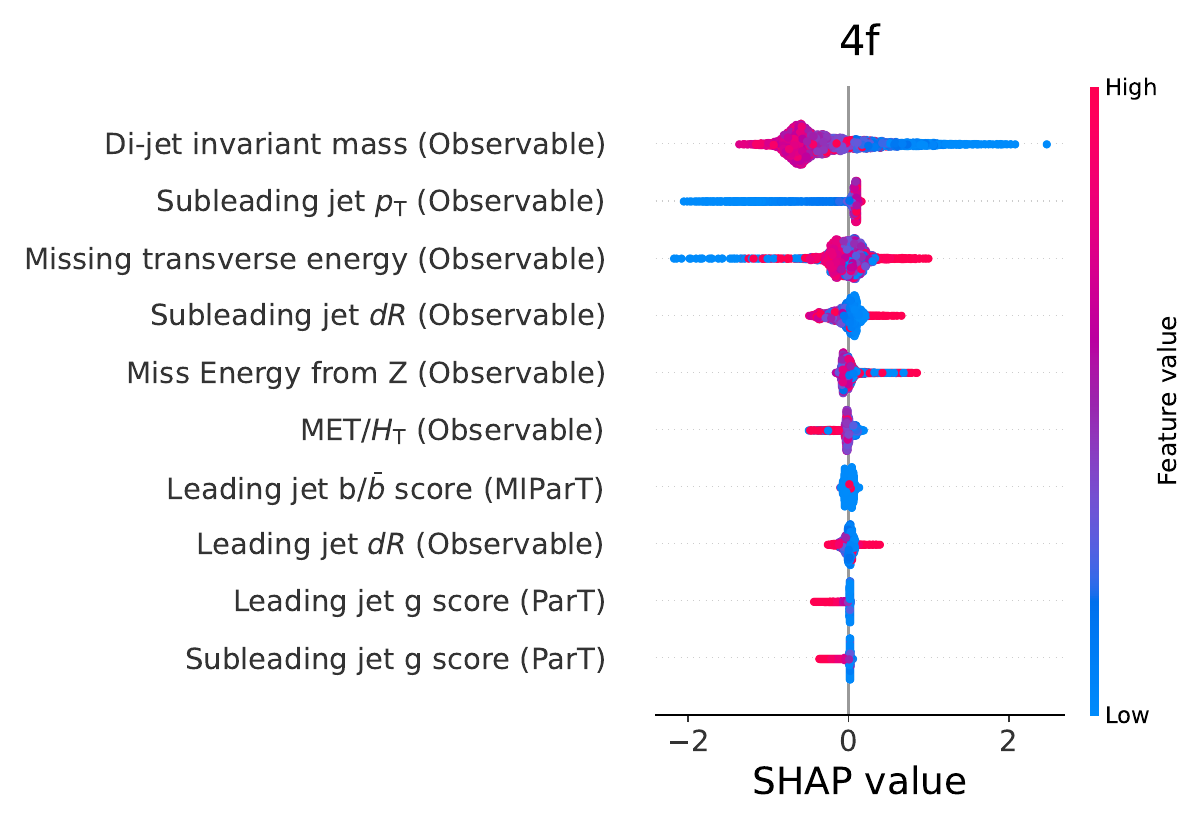} \\
        \textbf{(e)} & \textbf{(f)} \\
    \end{tabular}
}

\caption{
      SHAP summary plots for the event-level \texttt{XGB\_Combined} classifier, showing the ten most influential variables for each of the six output classes. In each panel, every point corresponds to a single event; the horizontal axis gives the SHAP value for that feature and class, while the colour encodes the feature value (from low in blue to high in red).
  Panels correspond to
  (a)~$H\to b\bar b$,
  (b)~$H\to c\bar c$,
  (c)~$H\to gg$,
  (d)~$H\to s\bar s$,
  (e)~$2f$,
  (f)~$4f$.
}
\label{fig:xgb_shap_multiclass}
\end{figure*}

While the gain ranking tells us which features matter on
average, it does not show how a given variable affects the
classification of individual events.  For that purpose we use SHAP
values~\cite{lundberg2017unifiedapproachinterpretingmodel}, which
provide a local, event--by--event decomposition of the classifier
output.

For a fixed output class $c$ (e.g., $H\to b\bar b$ or $2f$) and a given event $x$, the SHAP value of a feature $x_i$ represents its contribution to the classifier score $f_c(x)$, relative to the average score $\langle f_c \rangle$ over the event sample. 
In the XGBoost implementation used here, the classifier assigns a continuous score to each class, and the SHAP values approximately sum to the difference between this score and its average:
\begin{equation}
  f_c(x) - \langle f_c \rangle \;\simeq\; \sum_i \mathrm{SHAP}_{c,i}(x)\,.
\end{equation}
where the sum runs over all input features \(x_i\) in the event.

A positive SHAP value for class $c$ indicates that, given the observed value of this feature, the classifier is pushed to favour class $c$ more than it would for a typical event; a negative value means that the feature pushes the prediction away from that class.  The magnitude of the SHAP value indicates the strength of this effect.

Fig.~\ref{fig:xgb_shap_multiclass} shows SHAP ``summary'' plots for each of the six output classes.  In each panel, the vertical axis lists the 10 most important variables for that class.  Every point corresponds to one event in the test sample.  The horizontal position of a point is the SHAP value for that event and feature, so the spread along the $x$--axis indicates how strongly that variable can move events towards or away from the class under consideration.  The colour encodes the value of the feature itself: red points correspond to events where this feature takes a relatively large value (for example, a high $b$--tag score or a large MET), while blue points correspond to small values.  Regions where many points accumulate indicate that a given combination of feature value and SHAP impact is common in the sample; sparser regions typically correspond to rarer, more extreme events.

With this in mind, the structure of the SHAP plots is readily interpreted in physical terms.  
\begin{itemize}

\item For the $H\to b\bar b$ and $H\to c\bar c$ classes, shown in panels (a) and (b), the leading $b$- and $c$-jet scores show red points (high flavor score) predominantly at positive SHAP values and blue points (low score) at negative SHAP values.  This means that jets with a large $b$- or $c$-tagging probability systematically push the classifier towards the corresponding Higgs decay hypothesis, while jets with small scores make the event look less like $H\to b\bar b$ or $H\to c\bar c$.  This behaviour is exactly what one expects from a physically sensible heavy flavor tagger.

\item For the $H\to gg$ and $H\to s\bar s$ classes, shown in panels (c) and (d), the most important variables are the gluon and strange scores of both jets and the dijet invariant mass $m_{jj}$.  Large gluon (strange) scores favour the $H\to gg$ ($H\to s\bar s$) hypothesis, while small scores have the opposite effect.  
The fact that both red and blue points appear on either side of the vertical axis reflects the presence of correlations: for example, an event with a moderately high strange score may still be classified as background if other features (such as MET or jet kinematics) look incompatible with a $ZH(\nu\bar\nu)$ topology. 

\item For the $2f$ and $4f$ background classes, shown in panels (e) and (f), the most influential variables are global observables such as MET, $m_{jj}$, jet MET angular correlations and jet shape variables. Here the SHAP plots show that small MET values tends to yield positive SHAP values for the $2f$ class, indicating that such events look more like generic $e^+e^-\to q\bar q$ production than like a signal $ZH$ event.  
\end{itemize}

Overall, the SHAP analysis confirms and refines the picture obtained from the gain ranking: the classifier relies mainly on heavy and light flavor tagging information to distinguish between different Higgs decay modes, while global kinematic observables associated with the $ZH(\nu\bar\nu)$ topology are used to separate signal from the inclusive $2f/4f$ backgrounds.  The decision boundaries learned by the model are therefore well aligned with standard physics intuition.

\subsubsection*{Performance studies guided by interpretability}

The interpretability studies presented above naturally motivate targeted performance cross-checks. Here we investigate how the classifier performance changes when the input feature set is systematically restricted, guided by the global feature-importance and per-class SHAP analyses. We consider two complementary variations of the input configuration:

\begin{itemize}
  \item \textbf{Impact of individual jet taggers.}
  To probe the relative importance of the three jet taggers, we retrain the event-level classifier using only the ParT, MIParT, or PN flavor scores as jet inputs, while keeping the same set of event-level kinematic observables. The resulting precisions on $\sigma(ZH)\times\mathrm{Br}(H\to X)$ are summarised in Table~\ref{tab:bdt_part_comparison}. Using only the ParT scores already reproduces nearly the full performance of the \texttt{XGB\_Combined} configuration, consistent with the dominance of ParT variables observed in Fig.~\ref{fig:xgb_top20_importance}. Including PN and MIParT scores leads to small but non-negligible improvements, particularly for the $H\to gg$ and $H\to s\bar s$ channels, and reduces the spread among independent training runs. This indicates that the three taggers capture partially complementary aspects of the jet substructure.

    \item \textbf{Reduced-input study based on top features.}
    To assess how much of the classification performance can be retained with a smaller and more interpretable input space, we construct a reduced-input classifier, denoted \texttt{XGB\_Topfeatures}.
    The input variables of this classifier are selected in two steps.
    First, we retain the ten most important features according to the global feature-importance ranking obtained from the full \texttt{XGB\_Combined} model.
    Second, for each output class of the multi-class classifier, we add the five features with the largest absolute SHAP values in the corresponding class-specific SHAP summary.
    After removing duplicates between classes and with the globally selected features, the final input set contains 28 variables, listed in Table~\ref{tab:BDT_topfeatures_flat}.
    The \texttt{XGB\_Topfeatures} classifier is trained with the same architecture and training configuration as \texttt{XGB\_Combined}.
    The expected precisions obtained with \texttt{XGB\_Topfeatures}, reported in Table~\ref{tab:bdt_part_comparison}, are comparable to those of the full-feature configuration.
    This indicates that a large fraction of the discriminating power of the original classifier is already captured by a relatively small subset of physically interpretable features.

\end{itemize}

\begin{table*}[htbp]
\centering
\caption{ Comparison between the full XGBoost configuration (\texttt{XGB\_Combined}) and reduced configurations in which the jet inputs are restricted to only the ParT, MIParT or PN scores, while retaining the full set of event-level kinematic observables. Also shown is a reduced-input setup (\texttt{XGB\_Topfeatures}) guided by the feature importance and SHAP analyses, in which the overall input set is restricted to a compact subset of the most relevant variables. }
\label{tab:bdt_part_comparison}
\vspace{0.2cm}
\resizebox{\textwidth}{!}{%
\begin{tabular}{c c c c c c}
\toprule
\textbf{Channel} & \textbf{XGB\_Combined} & \textbf{XGB\_ParT} & \textbf{XGB\_MIParT} & \textbf{XGB\_PN} & \textbf{XGB\_Topfeatures} \\
\midrule
$Z \to \nu\bar{\nu},~ H \to b\bar{b}$ & 0.17\% & 0.17\% & 0.17\% & 0.18\% & 0.18\% \\
$Z \to \nu\bar{\nu},~ H \to c\bar{c}$ & 1.06\% & 1.06\% & 1.06\% & 1.07\% & 1.10\% \\
$Z \to \nu\bar{\nu},~ H \to gg$ & 0.50\% & 0.50\% & 0.51\% & 0.51\% & 0.54\% \\
$Z \to \nu\bar{\nu},~ H \to s\bar{s}$ & 68\% & 69\% & 69\% & 70\% & 73\% \\
\bottomrule
\end{tabular}
}
\end{table*}

\begin{table*}[htbp]
    \centering
    \caption{
      List of input variables used in the reduced-input XGBoost classifier (\texttt{XGB\_Topfeatures}). 
      Event-level and jet-level kinematic quantities are shown first, followed by jet-flavour tagging scores 
      from the ParT, PN and MIParT taggers for the leading and subleading jets.
      Here and in the following, $j_1$ and $j_2$ denote the leading and subleading jets in $p_\mathrm{T}$, respectively.
    }
    \label{tab:BDT_topfeatures_flat}
    \setlength{\tabcolsep}{8pt}
    \vspace{0.2cm}
    \begin{tabular}{@{\hspace{2pt}}p{6.5cm}p{8.0cm}@{\hspace{2pt}}}
        \toprule
        \textbf{Feature(s)} & \textbf{Description} \\
        \midrule
        \multicolumn{2}{c}{\textit{Event- and jet-level kinematics}} \\
        \midrule
        $\mathrm{MET}$ &
        Magnitude of the missing transverse energy \\
        $\mathrm{MET}/H_\mathrm{T}$ &
        MET to $H_\mathrm{T}$ ratio \\
        $\mathrm{MEZ}$ &
        The longitudinal components of the missing energy \\
        $m_{jj}$ &
        Invariant mass of the dijet system \\
        $\Delta\phi_{jj}$ &
        $\Delta\phi$ between the two jets \\
        $\Delta\phi(j_1,\mathrm{MET})$, $\Delta\phi(j_2,\mathrm{MET})$ &
        $\Delta\phi$ between Leading/subleading jet and MET \\
        $\Delta R(j_1,j_2)$ &
        Angular distance between the two jets \\
        $p_\textup{T}(j_2)$ &
        Transverse momenta of subleading jet \\
        \midrule
        \multicolumn{2}{c}{\textit{Jet-flavour tagging scores}} \\
        \midrule
        ParT (leading jet): $\bar b, \bar c, \bar s, s, g$ &
        ParT jet flavour scores for leading jet \\
        ParT (subleading jet): $\bar b, b, \bar c, c, \bar s, s, u, g$ &
        ParT jet flavour scores for subleading jet \\
        PN (leading jet): $\bar c, g$ &
        ParticleNet jet flavour scores for leading jet \\
        PN (subleading jet): $b, g$ &
        ParticleNet jet flavour scores for subleading jet \\
        MIParT (subleading jet): $b, d$ &
        MIParT jet flavour scores for subleading jet \\
        \bottomrule
    \end{tabular}
\end{table*}

\section{Comparison with holistic event-level classification}
\label{app:holistic}

Reference~\cite{Zhu:2025eoe} has proposed a `holistic' strategy for classifying hadronic Higgs decays at lepton colliders.  In that approach, the entire reconstructed event is treated as a single particle cloud and passed to an event-level ParticleNet classifier. All reconstructed particles and their low-level features (kinematics, particle identification, track information, etc.) enter directly as inputs, without an explicit intermediate jet reconstruction or jet-by-jet flavor tagging step. The network is therefore responsible for learning both the internal jet substructure and the global event topology in a single stage and produces, for each event, likelihoods for a set of signal hypotheses.

By contrast, the strategy developed in this work decomposes the problem into two physically motivated levels:  
(i)~jet-level flavor classification using ParticleNet, ParT and MIParT, and  
(ii)~event-level classification using XGBoost, which takes as inputs both global kinematic observables and the jet-flavor probabilities from the three taggers.
This modular structure explicitly separates microscopic jet-substructure information from global event kinematics, which is advantageous for interpretability and for the eventual treatment of systematic uncertainties.

To compare the physics reach of the two strategies in a controlled way, we reproduce, as closely as possible, the setup of Ref.~\cite{Zhu:2025eoe}. In particular:
\begin{itemize}
  \item We restrict the comparison to the $e^+e^-\to ZH$ channel with
        $Z\to\nu\bar\nu$;
  \item We consider only the four hadronic Higgs decay modes
        $H\to b\bar b$, $c\bar c$, $gg$ and $s\bar s$, treating them as
        mutually exclusive classes; and
  \item We do not include the $2f$ and $4f$ backgrounds in this
        comparison, in order to focus on the relative separation power
        among the different Higgs decay modes themselves.
\end{itemize}
Using exactly the same significance definition and optimisation procedure as in Sec.~\ref{sec:results_sens}, we determine the expected relative precision on $\sigma(ZH)\times\mathrm{Br}(H\to X)$ for each channel in our two-stage framework and normalise the results to an integrated luminosity of $20~\text{ab}^{-1}$. The corresponding numbers from the holistic ParticleNet study are taken from Ref.~\cite{Zhu:2025eoe}. The comparison is summarised in Table~\ref{tab:holistic_compare}.

\begin{table*}[htbp]
    \centering
    \caption{
        Comparison of the relative statistical precision on
        $\sigma(ZH)\times\mathrm{Br}(H\to X)$ in the
        $Z\to\nu\bar\nu$ channel between this work (two-stage 
        classification) and the holistic ParticleNet event-level
        classifier of Ref.~\cite{Zhu:2025eoe}.  Only the four Higgs
        decay modes $H\to b\bar b$, $c\bar c$, $gg$ and $s\bar s$ are
        considered, without including the $2f$ and $4f$ backgrounds.
        All values are normalised to an integrated luminosity of
        $20~\text{ab}^{-1}$.
    }
    \label{tab:holistic_compare}
    \vspace{0.2cm}
    \setlength{\tabcolsep}{10pt}
    \begin{tabular}{c c c}
        \toprule
        \textbf{Channel} 
            & \textbf{This work} 
            & \textbf{Holistic} \\
        \midrule
        $Z\to\nu\bar\nu,\; H\to b\bar b$ 
            & 0.17\% & 0.14\% \\
        $Z\to\nu\bar\nu,\; H\to c\bar c$ 
            & 0.78\% & 0.72\% \\
        $Z\to\nu\bar\nu,\; H\to gg$ 
            & 0.46\% & 0.46\% \\
        $Z\to\nu\bar\nu,\; H\to s\bar s$ 
            & 23\% & 29\% \\
        \bottomrule
    \end{tabular}
    \vspace{0.5em}
\end{table*}

The two approaches yield very similar sensitivities. For the high-rate channels $H\to b\bar b$ and $c\bar c$, the relative precisions agree within a few hundredths of a percent, well below the level at which other theoretical or experimental systematics are expected to enter. In the statistically more challenging $H\to s\bar s$ channel, both methods achieve relative precisions at the level of $20$--$30\%$ under the idealised assumptions of this comparison. 
These results indicate that, at least in this simplified setup, the explicit use of jet reconstruction and dedicated jet-flavor taggers does not lead to any loss of information compared to a fully holistic treatment of the event; both strategies appear to saturate the discriminating power available in the simulated samples.

However, it is important to emphasise the conceptual differences between the two approaches, which become particularly relevant once systematic effects and more complex final states are considered:
\begin{itemize}
  \item \textbf{Modularity and calibration.}
        In the two-stage framework, the jet-level taggers and the
        event-level classifier are separate components.
        The jet taggers can in principle be calibrated and validated
        using control samples enriched in specific flavours
        (e.g.\ $Z\to b\bar b$, $Z\to c\bar c$, light-flavor and gluon
        jets), while the event-level XGBoost can be retrained or fine-tuned
        using these calibrated inputs.
        This modularity is advantageous for controlling systematic
        uncertainties in a real experiment.
        In a holistic network, the flavor and topology information are
        entangled in a single, monolithic model, making it more
        challenging to isolate and correct potential mismodellings in
        specific physics components.

  \item \textbf{Reusability and flexibility.}
        Once trained, jet-level taggers can be reused across many
        different analyses and channels (e.g.\ other Higgs production
        modes or $Z$-pole measurements), with only the event-level
        classifier needing to be adapted to a new topology.
        A holistic event-level network would typically have to be
        retrained for each new physics channel or detector configuration,
        which is computationally more demanding and less transparent.

  \item \textbf{Interpretability.}
        The two-stage method provides a clear physics picture of how
        different ingredients contribute: heavy-flavor scores drive the
        separation of $H\to b\bar b$ and $H\to c\bar c$, strange and
        gluon scores control the $H\to s\bar s$ versus $H\to gg$
        discrimination, and global kinematics (such as missing energy
        and jet angular correlations) distinguish $ZH(\nu\bar\nu)$ from
        generic $2f/4f$ backgrounds.
        In a holistic model this decomposition is implicit and must be
        inferred a posteriori, e.g.\ via saliency maps, which can be
        more difficult to relate to specific experimental systematics.
\end{itemize}

In summary, the quantitative comparison in Table~\ref{tab:holistic_compare} shows that our two-stage jet-plus-event classifier achieves a statistical performance that is fully comparable to that of the holistic ParticleNet approach under the same simplified benchmark. At the same time, the modular structure of the two-stage strategy provides practical advantages for calibration, systematic-uncertainty control and reuse across channels, which are essential considerations for precision Higgs measurements at future lepton colliders.

\clearpage

\bibliographystyle{JHEP}
\bibliography{apssamp.bib}

\end{document}